\documentclass[]{iopart}

\usepackage{amssymb}
\usepackage{graphicx}
\usepackage{cite}
\usepackage[colorlinks=true,linkcolor=blue,citecolor=blue ,urlcolor=blue]{hyperref}

\begin{document}

\newcommand{\nota}[1]{\textcolor[rgb]{1,0,0}{{#1}}}

\title[Symmetry Adapted Coherent States for Three-Level Atoms]%
{Symmetry Adapted Coherent States for Three-Level Atoms Interacting
with One-Mode Radiation}

\author{R. L\'opez--Pe\~na, S. Cordero, E. Nahmad--Achar and \\ O. Casta\~nos}

\address{%
Instituto de Ciencias Nucleares, Universidad Nacional Aut\'onoma de
M\'exico, Apartado Postal 70-543, 04510
M\'exico DF,  M\'exico}

\ead{\mailto{lopez@nucleares.unam.mx}}

\begin{abstract}
We introduce a combination of coherent states as variational
test functions for the atomic and radiation sectors to describe a system of $N_{a}$ three-level atoms interacting with a one-mode quantised
electromagnetic field, with and without the rotating wave
approximation, which preserves the symmetry presented by the Hamiltonian. These provide us with the possibility of finding analytical 
solutions for the ground and first excited states. 
We study the properties of these solutions for the $V$-configuration in the double
resonance condition, and calculate the expectation values of the number of photons,
the atomic populations, the total number of excitations,
and their corresponding fluctuations. We also calculate the photon
number distribution and the linear entropy of the reduced density
matrix to estimate the entanglement between matter and radiation. For the first time, we exhibit analytical expressions for all of these quantities, as well as an analytical description for the phase diagram in parameter space, which distinguishes the normal and collective regions, and which gives us all the quantum phase transitions of the ground state from one region to the other as we vary the interaction parameters (the matter-field coupling constants) of the model, in functional form.
\end{abstract}

\maketitle

\section{Introduction}

Two-level systems have been extensively studied in quantum optics~\cite{garraway2011,nahmad2012}. The promise that qutrits and qudits in general can extend the possibilities of $2$-level systems in quantum information and other scenarios makes the study of higher-dimensional  quantum systems, in particulat $3$- and $4$-level systems, interesting.

Implementations of qutrit channels have been demonstrated where two photon-polarization-qubits form a biphotonic qutrit~\cite{lanyon2008} and, though difficult in practice, biphoton-photon entanglement has proved manageable~\cite{imre2013}.

There has been detailed research on the physical phenomena involving
two-photon processes in one three-level
atom~\cite{yoo1985,li1985,peng1992}. More recently, there has been
interest in the phase states of a three-level atom interacting through
one and two modes of radiation~\cite{aliskenderov1991, klimov2003}.
Phase transitions in two-color superradiance has been discussed
in~\cite{hayn2011} for the $\Lambda$-configuration.  The influence of
a Kerr-like medium on the temporal evolution of the second-order
correlation function for a $3$-level atom has been studied in all
configurations~\cite{abdel2007}.  The quantum phase diagrams of
three-level atoms interacting with a one-mode radiation field have
been obtained analytically, for all the configurations and in the
rotating wave approximation (RWA) in~\cite{cordero2013a,cordero2013b}.

In this paper we review the calculation of the energy surface of
$3$-level atoms interacting with a one mode radiation field, with and
without the RWA approximation. The energy surface is defined by the
expectation value of the Hamiltonian with respect to the product
of Weyl-Heisenberg coherent states for the field and the totally
symmetric U($3$) coherent states for the matter. The analytical form
of these energy surfaces allows us to:
\begin{description}
\item[{\it i)}] extend the quantum phase diagrams obtained with the
  RWA for the transition from the normal to the superradiant regimes
  of the atoms to the full Hamiltonian. 
\item[{\it ii)}] construct  {\sl symmetry-adapted states} that give a better
  description of the ground state of the system, and additionally to
  have an approximation to the first excited state. 
\end{description}

These new states allow to properly determine the entanglement between
matter and radiation, and the statistical behaviour of the total
number of excitations operator.  In section 2 we establish the
Hamiltonian of the system in terms of bosonic operators for the field
and U($3$) generators for the matter. We determine the energy surface,
with and without the RWA approximation, calculating the expectation
value of the corresponding Hamiltonians with respect to matter and
field coherent states, and we describe how to use it as an
approximation of the ground state of the system, in section
3. Establishing a relation between both energy surfaces allows us to
extend the analytical results of the quantum phase diagrams obtained
in~\cite{cordero2013a, cordero2013b}.  In section 4 we give the
minima of the full Hamiltonian energy surface (with respect to the
standard product of coherent states) for the $V$-configuration under
the double resonance condition. These minima are substituted into the
general expressions for the energy surface and other observables. As
an example we present the results obtained for the mean square
deviation of the number of photons.  In section 5 we show that the
model Hamiltonian in the RWA approximation has the total number of
excitations operator $\mathbf{M}$ as a constant of motion while the
full Hamiltonian has only the parity operator $e^{i \,\pi\,\mathbf{M}}$ as a
constant of motion.  These results lead us to introduce the
symmetry-adapted coherent states (SACS) for the model without the RWA
approximation.  At the end of the section we give the corresponding
energy surfaces of the model Hamiltonian for the normal and collective
regimes, and for the even and odd parity cases.  In section 6 we
compare the results obtained for the SACS energy surface with the
coherent energy surface. The same is done for other matter and field
observables as well as for the photon number distribution. In
particular, our methodology allows us to calculate the matter-field
entanglement of the system for the SACS and compare the result
with the linear entropy for the coherent states. Finally we give a
summary of the obtained results.

\section{The Model}

We consider a system composed of $N_{a}$ three-level atoms interacting
with a one-mode of quantised electromagnetic field in a cavity. The
interaction occurs only through the atomic electric dipole moment,
i.e., the electric quadrupole and magnetic dipole interactions are
smaller by a factor of the fine structure constant $\alpha\approx
1/137$, and are thus neglected. We also use the long-wavelength
approximation, i.e., the size of an atom is much smaller than the
wavelength of the electromagnetic radiation. Then the full Hamiltonian
model for this system is given by~\cite{yoo1985}
  \begin{equation}
      \mathbf{H}=\Omega\,\mathbf{a}^{\dagger}\mathbf{a}+\sum_{j=1}^{3}
      \,\omega_{j}\,\mathbf{A}_{jj}-
      \sum_{j<k}^{3}\frac{\mu_{jk}}{\sqrt{N_{a}}}
      \left(\mathbf{a}^{\dagger}+\mathbf{a}\right)\left(\mathbf{A}_{jk}
      +\mathbf{A}_{kj}\right)\ .
      \label{eq01}
   \end{equation}
In this equation, $\mathbf{a}^{\dagger}$ and $\mathbf{a}$ denote the
creation and annihilation operators of the one-mode radiation field of
frequency $\Omega$, $\mathbf{A}_{ij}$ denotes the collective matter
operators defined by
$\mathbf{A}_{ij}:=\sum_{r=1}^{N_{a}}\mathbf{A}_{ij}^{(r)}$, where
$\mathbf{A}_{ij}^{(r)}$ stands for the atomic operator which changes
the atom $r$ from level $j$ to level $i$, the atomic levels are
ordered such that $\omega_{1}\leq\omega_{2}\leq\omega_{3}$, and
$\mu_{ij}$ denote the dipolar intensities. A particular atomic
configuration is set by properly choosing one dipolar parameter
$\mu_{i j}=0$. This Hamiltonian describes a dilute gas whose atoms
only interact through the radiation field.
   
The collective operators $\mathbf{A}_{ij}$ satisfy the
commutation relations of the $U(3)$ Lie algebra
   \begin{equation}
      \Big[\mathbf{A}_{ij},\,\mathbf{A}_{kl}\Big]
      =\delta_{jk}\,\mathbf{A}_{il}-\delta_{il}\,
      \mathbf{A}_{kj}\ ,
   \end{equation}
and the linear and quadratic invariants are given by
   \begin{equation}
      \mathbf{N_a} = \sum_{k=1}^{3}\mathbf{A}_{kk} \, , \qquad
      \mathbf{N^2_a} + 2 \, \mathbf{N_a} =
      \sum_{k,j=1}^{3}\mathbf{A}_{kj} \ \mathbf{A}_{jk} \, ,
   \end{equation}
where $N_a$ denotes the number of atoms of the system.

\subsection{Symmetric coherent states for the matter}

If we consider a system of identical atoms, we can represent
the collective atomic operators $\mathbf{A}_{ij}$ in the form
   \begin{equation}
      \mathbf{A}_{ij}=\mathbf{b}_{i}^{\dagger}\,\mathbf{b}_{j}\ ,
   \end{equation}
where the creation $\mathbf{b}^{\dagger}_{i}$ and annihilation $\mathbf{b}_{j}$
operators with $i,\,j=1,2,3$ satisfy the commutation relations
   \begin{equation}
      \fl
      \left[\mathbf{b}_{j},\,\mathbf{b}_{k}^{\dagger}\right]=\delta_{jk}\,\mathbf{1}\ ,
      \quad\left[\mathbf{b}_{j},\,\mathbf{b}_{k}\right]=0\ ,
      \quad\left[\mathbf{b}_{j}^{\dagger},\,\mathbf{b}_{k}^{\dagger}\right]=0\ ,
      \quad j,\,k=1,\,2,\,3\ .
   \end{equation}

If we define the operators
   \begin{equation}
      \fl
      \mathbf{\Gamma}^{\dagger}(\gamma):=\frac{1}{\sqrt{\sum_{k}|\gamma_{k}|^{2}}}
      \sum_{j=1}^{3}\,\gamma_{j}\,\mathbf{b}_{j}^{\dagger}\ ,\quad
      \mathbf{\Gamma}(\gamma):=\frac{1}{\sqrt{\sum_{k}|\gamma_{k}|^{2}}}
      \sum_{j=1}^{3}\,\gamma_{j}^{\ast}\,\mathbf{b}_{j}\ ,
   \end{equation}
which satisfy the commutation relations of boson operators, then it is straightforward
to define the condensate or totally symmetric $U(3)$-coherent state of $N_{a}$ atoms
as follows:
   \begin{eqnarray}
      \vert N_{a};\,\gamma\rangle&:=&\frac{1}{\sqrt{N_{a}!}}\,\left[
      \mathbf{\Gamma}^{\dagger}(\gamma)\right]^{N_{a}}\,\vert 0\rangle
      =\frac{1}{(\sum_{k}|\gamma_{k}|^{2})^{N_{a}/2}}\vert N_{a};\,\gamma\}\ ,
   \end{eqnarray}
where in the last expression we defined the unnormalised
coherent state which can be written as~\cite{gilmore1975}
   \begin{eqnarray}
      \vert N_{a};\,\gamma\}
      &=&\sum_{n_{1}+n_{2}+n_{3}=N_{a}}
      \sqrt{\frac{N_{a}!}{n_{1}!\,n_{2}!\,n_{3}!}}\,\gamma_{1}^{n_{1}}
      \,\gamma_{2}^{n_{2}}\,\gamma_{3}^{n_{3}}\,
      \vert n_{1},\,n_{2},\,n_{3}\rangle\ ,\nonumber
   \end{eqnarray}
and where $\vert n_{1},\,n_{2},\,n_{3}\rangle$ denotes the state with
$n_{i}$ atoms in levels $i=1,\,2,\,3$. In these
expressions $\gamma$ stands for
$(\gamma_{1},\,\gamma_{2},\,\gamma_{3})$.

The scalar product of these states is
   \begin{equation}
      \{N_{a};\,\gamma\vert N_{a}^{\prime};\,\gamma^{\prime}\}
      =\delta_{N_{a}\,N_{a}^{\prime}}\,\left(\gamma^{\ast}
      \cdot\gamma^{\prime}\right)^{N_{a}}\ ,
   \end{equation}
with
   \[
      \gamma^{\ast}\cdot\gamma^{\prime}=\sum_{j=1}^{3}\,
      \gamma_{j}^{\ast}\,\gamma^{\prime}_{j}\ .
   \]

The $U(3)$ Lie algebra generators can be represented in this basis by
the differential operators
   \begin{equation}
      \mathbf{A}_{jk}:=\mathbf{b}_{j}^{\dagger}\mathbf{b}_{k}\longmapsto
      \gamma_{j}^{\ast}\,\frac{\partial\phantom{\gamma_{j}^{\ast}}}{
      \gamma_{k}^{\ast}}\ .
   \end{equation}
This result allows us to obtain the matrix elements of the generators
$A_{jk}$ in the unnormalised coherent states:
   \begin{equation}
   \fl 
      \{N_{a};\,\gamma\vert \mathbf{A}_{jk}\vert N_{a}^{\prime};\,\gamma^{\prime}\}
      =N_{a}\,\frac{\gamma_{j}^{\ast}\,\gamma^{\prime}_{k}}{
      \gamma^{\ast}\cdot\gamma^{\prime}}\,
      \{N_{a};\,\gamma\vert N_{a}^{\prime};\,\gamma^{\prime}\}\ ,\quad
      \gamma^{\ast}\cdot\gamma^{\prime}\not=0\ .
   \end{equation}
The matrix elements of operator $A_{ij}A_{kl}$ can also be calculated
using this method and the result is given by
   \begin{eqnarray}
      \fl
      \{N_{a};\,\gamma\vert\,\mathbf{A}_{ij}\,\mathbf{A}_{kl}
      \,\vert\,N_{a}^{\prime};\,\gamma^{\prime}\}&=&\frac{N_a \, \gamma_{i}^{\ast}\,\gamma^{\prime}_{l}}{\gamma^{\ast}
      \cdot\gamma^{\prime}}\Big[\,(N_{a}-1)
      \,\frac{\gamma^{\prime}_{j}\,\gamma_{k}^{\ast}}{
      (\gamma^{\ast}\cdot\gamma^{\prime})}
      +\delta_{jk}\,
      \Big]\,\{N_{a};\,\gamma\vert\,N_{a}^{\prime};\,\gamma^{\prime}\}\ .
   \end{eqnarray}

In the normalised coherent states we can divide numerator and
denominator by $\gamma_{1}$, and this quantity will not appear in the
final expressions. The same can be done for the matrix elements of the
$U(3)$ generators between coherent states. This shows that the
introduced coherent states actually depend only on two complex
variables instead of three. We will keep the redundant expression
$\gamma=(\gamma_{1},\,\gamma_{2},\,\gamma_{3})$ to simplify the
notation.  However, we must set $\gamma_{1}=1$. Also, because atomic
coherent states with different number of atoms are orthogonal, we will
consider always the same number of atoms $N_{a}$ and drop this number
from the labelling of the states.

\section{Energy Surface}

In order to study the ground state of the Hamiltonian~(\ref{eq01}), we
calculate its energy surface which is defined as the expectation value
with respect to
	\begin{equation}
		\vert\alpha;\,\gamma\}:=
		\vert\alpha\}\otimes\vert\,\gamma\}\ .
	\end{equation}
Here we use also the unnormalised Weyl--Heisenberg coherent states
	\begin{equation}
		\vert\alpha\}:=\exp\left(\alpha\,\mathbf{a}^{\dagger}\right)\,\vert 0\rangle\ .
	\end{equation}
Thus, the energy surface has the form
	\begin{eqnarray}
		\fl
		&&E(\alpha,\,\gamma)
		:=\{\alpha;\,\gamma\vert\,\mathbf{H}\,
		\vert\alpha;\,\gamma_{2},\,\gamma_{3}\}/
		\{\alpha;\,\gamma\vert
		\alpha;\,\gamma\}\nonumber\\
		\fl
		&&\ =\Omega\,\left|\alpha\right|^{2}+
		\frac{1}{\gamma^{\ast}\cdot\gamma}\Big\{\,N_{a}\,
		\sum_{i=1}^{3}\omega_{i}\,\left|\gamma_{i}\right|^{2}
		-\sqrt{N_{a}}\,\sum_{i<j}^{3}\mu_{ij}\,\left(
		\gamma_{i}^{\ast}\gamma_{j}+\gamma_{j}^{\ast}\gamma_{i}
		\right)\,\left(\alpha^{\ast}+\alpha\right)\Big\}\ .
	\end{eqnarray}
Using the polar form of the complex numbers
	\begin{equation}
		\alpha:=\varrho\,\exp(i\,\varphi)\ ,\quad
		\gamma_{j}:=\varrho_{j}\,\exp(i\,\varphi_{j})\ ,\
		j=1,\,2,\,3\ ,
           \label{eq011}
	\end{equation}
and setting $\gamma_{1}=1$, this function takes the form
	\begin{eqnarray}
		\fl
		&&E(\varrho,\,\varphi,\,\varrho_{j},\,\varphi_{j})
                =\Omega^{2}\,\varrho^{2}+\Big\{N_{a}\,\Big[\omega_{1}
		+\omega_{2}\,\varrho_{2}^{2}+\omega_{3}\,\varrho_{3}^{2}
		\Big]-4\,\sqrt{N_{a}}\Big[\,\mu_{12}\,\varrho_{2}\,\cos\varphi_{2}
		\nonumber\\
		\fl
		&&\quad+\mu_{13}\,\varrho_{3}\,\cos\varphi_{3}
		+\mu_{23}\,\varrho_{2}\,\varrho_{2}\,
		\cos(\varphi_{2}-\varphi_{3})\Big]\,
		\varrho\,\cos\varphi\Big\}/\left(1+\varrho_{2}^{2}
		+\varrho_{3}^{2}\right)\ .
	\end{eqnarray}
If we assume that the interaction intensities $\mu_{ij}$ are
non-negative numbers, the minimum value of this energy surface is
obtained when the angles satisfy the conditions
	\begin{equation}
		\cos\varphi_{2}\,\cos\varphi=\cos\varphi_{3}\,\cos\varphi
		=\cos(\varphi_{2}-\varphi_{3})\,\cos\varphi\equiv 1\ ,
	\end{equation}
because in this case the contribution of the atom-field interaction
Hamiltonian diminishes the value of the energy surface. Actually,
this results from considering the hessian of the Hamiltonian; in any case
the product of these cosines times the dipolar intensity parameter must be
positive in order to have a minimum. Thus
	\begin{eqnarray}
		\fl
		&&E_{min}(\varrho_{c},\,\varrho_{2\,c},\,\varrho_{3\,c})
		=\Omega\,\varrho_{c}^{2}+\Big\{N_{a}\,\Big[\omega_{1}
		+\omega_{2}\,\varrho_{2\,c}^{2}+\omega_{3}\,\varrho_{3\,c}^{2}
		\Big]\nonumber\\
		\fl
		&&\qquad \qquad \quad -4\,  \sqrt{N_{a}} \,\varrho_{c}\Big[\mu_{12}\,\varrho_{2\,c}
		+\mu_{13}\,\varrho_{3\,c}
		+\mu_{23}\,\varrho_{2\,c}\,\varrho_{3\,c}\Big]\Big\}/
		\left(1+\varrho_{2\,c}^{2}+\varrho_{3\,c}^{2}\right)\, ,
		\label{emin}
	\end{eqnarray}
where $\varrho_{c}$, $\varrho_{2\,c}$ and $\varrho_{3\,c}$ denote the
minima critical values of the corresponding variables.

\subsection{Comparison with the RWA Hamiltonian}

The energy surface of the Hamiltonian in the rotating wave
approximation is given by
	\begin{equation}
		\fl
		E_{\hbox{\tiny{RWA}}}(\alpha,\,\gamma)
		=\Omega\left|\alpha\right|^{2}
		+\frac{1}{\gamma^{\ast}\cdot\gamma}\Big\{N_{a}
		\sum_{i=1}^{3}\omega_{i}\left|\gamma_{i}\right|^{2}
		-\sqrt{N_{a}}\sum_{i<j}^{3}\mu_{ij}\left(
		\gamma_{i}^{\ast}\,\gamma_{j}\,\alpha^{\ast}
		+\gamma_{j}^{\ast}\,\gamma_{i}\alpha\right)\Big\} .
	\end{equation}
Using polar coordinates~(\ref{eq011}) we get
	\begin{eqnarray}
		\fl
		&&E_{\hbox{\tiny{RWA}}}(\varrho,\,\varphi,\,\varrho_{j},\,\varphi_{j})
                =\Omega^{2}\,\varrho^{2}+\Big\{N_{a}\Big[\omega_{1}
		+\omega_{2}\,\varrho_{2}^{2}+\omega_{3}\,\varrho_{3}^{2}
		\Big]-2\,\sqrt{N_{a}}\,\varrho \Big[
		\mu_{12}\,\varrho_{2} \cos(\varphi_{2}-\varphi)
		\nonumber\\
		\fl
		&&\qquad+\mu_{13}\,\varrho_{3}\,\cos(\varphi_{3}-\varphi)
		+\mu_{23}\,\varrho_{2}\,\varrho_{2}\,
		\cos(\varphi_{3}-\varphi_{2}-\varphi)\Big]\,
		\Big\}/\left(1+\varrho_{2}^{2}
		+\varrho_{3}^{2}\right)\ .
	\end{eqnarray}
In this case the minimum value of the energy surface is found when
the angles satisfy the conditions
	\begin{equation}
		\cos(\varphi_{2}-\varphi)=\cos(\varphi_{3}-\varphi)
		=\cos(\varphi_{3}-\varphi_{2}-\varphi)\equiv 1\ ,
	\end{equation}
and its expression is
	\begin{eqnarray}
	\fl
		&&\left(E_{\hbox{\tiny{RWA}}}\right)_{min}(\varrho_{c},\,\varrho_{2\,c},\,\varrho_{3\,c})=
		\Omega\,\varrho_{c}^{2}+\Big\{N_{a}\,\Big[\omega_{1}
		+\omega_{2}\,\varrho_{2\,c}^{2}+\omega_{3}\,\varrho_{3\,c}^{2}
		\Big]\nonumber\\
		\fl
		&&\qquad-2\,\sqrt{N_{a}}\,\varrho\,\Big[\mu_{12}\,\varrho_{2\,c}
		+\mu_{13}\,\varrho_{3\,c}
		+\mu_{23}\,\varrho_{2\,c}\,\varrho_{3\,c}\Big]\Big\}/
		\left(1+\varrho_{2\,c}^{2}+\varrho_{3\,c}^{2}\right)\ .
	\end{eqnarray}

Comparing this equation with Eq.~(\ref{emin}) we note that the energy
surfaces $E_{min}$ and $(E_{\hbox{\tiny{RWA}}})_{min}$ coincide if we carry out
the following identification of the atom-field interaction parameters:
	\begin{equation}
		\left(\mu_{jk}\right)_{\hbox{\tiny{RWA}}}
		\longrightarrow 2\,\left(\mu_{jk}\right) \ .
		\label{eq23}
	\end{equation}
Said differently, $E_{\hbox{\tiny{RWA}}}$ will inherit the properties
of $E_{min}$ at values of $(\mu_{ij})_{\hbox{\tiny{RWA}}}$ equal to
$\frac{1}{2}\mu_{ij}$.

We want to emphasize that equation~(\ref{eq23}) is valid for all the
atomic configurations. Therefore the analytic expressions for the
quantum phase diagrams of the model Hamiltonian with the RWA
approximation can be now extended to the full Hamiltonian, i.e., in
equations (5)-(7) of~\cite{cordero2013a} or equations (36)-(38)
of~\cite{cordero2013b} we properly substitute the dipolar interactions
$\mu_{\hbox{\tiny{RWA}}} \to 2\, \mu$.

\section{Description of observables}

We are interested only in the ground state of the system, thus we will
consider only critical points whose hessian has positive eigenvalues.
For the normal regime, the minima critical points are given by
$\varrho_{c}=\varrho_{2\,c}=\varrho_{3\,c}=0$, while in the
superradiant regime there are not, in general, analytical expressions
for the minima, and the calculation must be done numerically. However,
for the $V$-configuration ($\mu_{23}=0$) in the double-resonance case,
i.e., $\omega_{2}=\omega_{3}$, there is an analytical formula for the
minima~\cite{cordero2013b}. In this case we have, in the collective
regime for the radiation variable,
   \begin{equation}
      \varrho_{c}=\frac{\sqrt{N_{a}}}{\Omega}\frac{
      \mu_{12}\,\varrho_{3c}+\mu_{13}\,\varrho_{3c}}{
      1+\varrho_{2c}^{2}+\varrho_{3c}^{2}}\ .
      \label{eq025}
   \end{equation}
For the atomic variables there are two solutions.  The critical point
$\varrho_{c}=\varrho_{2c}=\varrho_{3c}=0$ exists for all the values of
the parameter interaction strengths $\mu_{12}$ and $\mu_{13}$. However
it is not a minimum when
$\mu_{13}^{2}+\mu_{23}^{2}>\Omega\,\omega_{3}/4$. When this condition
is satisfied, which is called the collective regime, there is a
solution to the equations of the critical conditions which gives a
negative energy. In the collective regime, where there are many
configurations with different number of photons contributing to the
state, the minimum critical points are given by
   \begin{eqnarray}
      \varrho_{2c}&=&\mu_{12}\,\sqrt{\frac{\mu_{12}^{2}+\mu_{13}^{2}
      -\Omega\,\omega_{3}/4}{(\mu_{12}^{2}+\mu_{13}^{2})
      (\mu_{12}^{2}+\mu_{13}^{2}+\Omega\,\omega_{3}/4)}}\ ,
      \label{eq026}\\
      \varrho_{3c}&=&\mu_{13}\,\sqrt{\frac{\mu_{12}^{2}+\mu_{13}^{2}
      -\Omega\,\omega_{3}/4}{(\mu_{12}^{2}+\mu_{13}^{2})
      (\mu_{12}^{2}+\mu_{13}^{2}+\Omega\,\omega_{3}/4)}}\ .
      \label{eq027}
   \end{eqnarray}
When substituting these results into the expression for the energy
surface we obtain, in the collective regime,
   \begin{equation}
      E_{min}^{(V)}=-\frac{1}{\Omega}\frac{(\mu_{12}^{2}+\mu_{13}^{2}
      -\Omega\,\omega_{3}/4)^{2}}{\mu_{12}^{2}+\mu_{13}^{2}}\ ,
      \quad\mu_{13}^{2}+\mu_{23}^{2}>\Omega\,\omega_{3}/4\ ,
   \end{equation}
and $E_{min}^{(V)}=0$ when
$\mu_{13}^{2}+\mu_{23}^{2}\le\Omega\,\omega_{3}/4$ (normal regime).

If we calculate the expression for the mean square deviation of the
number of photons, which is equal to the expectation value of the
number of photons, we obtain
   \begin{equation}
      \fl
      (\Delta(\mathbf{a}^{\dagger}\mathbf{a})_{V})^{2}=\langle
      \mathbf{a}^{\dagger}\mathbf{a}\rangle_{V}=N_{a}\frac{
      (\mu_{12}^{2}+\mu_{13}^{2}-\Omega\,\omega_{3}/4)
      (\mu_{12}^{2}+\mu_{13}^{2}+\Omega\,\omega_{3}/4)}{
      \Omega^{2}\,\left(\mu_{12}^{2}+\mu_{13}^{2}\right)}\ ,
   \end{equation}
in the collective regime
($\mu_{12}^{2}+\mu_{13}^{2}>\Omega\,\omega_{3}/4$), and $0$ in the
normal regime ($\mu_{12}^{2}+\mu_{13}^{2}\le\Omega\,\omega_{3}/4$).

\section{Symmetry Adapted Coherent States}

In the previous section we have shown that there is a simple relation
between all the matter and field observables when the RWA
approximation is used and when it is not. Therefore, we can then extend
all the analytical expressions in the RWA approximation to the case of
the full Hamiltonian.

However, the solutions of the stationary Schr\"odinger equation in the
RWA approximation have an extra constant of motion besides the
energy: the total excitation number. This quantity depends on the
considered atomic configuration
	\begin{equation}
		\mathbf{M}=\mathbf{a}^{\dagger}\mathbf{a}
                +\lambda_{2}\,\mathbf{A}_{22}
		+\lambda_{3}\,\mathbf{A}_{33}\ ,
	\end{equation}
where the values of $\lambda_{j}$, $j=2,\,3$, are given in
Table~\ref{t1} for the three possible configurations.

   \begin{table}[h]
        \caption{Values of parameters $\lambda_{i}$ for each of the
        atomic configurations. For these values the operator
        $\mathbf{M}=\mathbf{a}^{\dagger}\mathbf{a}+\lambda_{2}\,\mathbf{A}_{22}
	+\lambda_{3}\,\mathbf{A}_{33}$ is a constant of motion of the system.}
	\begin{center}
	\begin{tabular}{c||c|c}
	Configuration&\phantom{mm}$\lambda_{2}$\phantom{mm}
        &\phantom{mm}$\lambda_{3}$\phantom{mm}\\[0.8ex]
	\hline\hline
	$\Xi$&1&2\\[0.8ex]
	\hline
	$\Lambda$&0&1\\[0.8ex]
	\hline
	$V$&1&1
	\end{tabular}
	\end{center}
	\label{t1}
   \end{table}

For the solutions without the RWA approximation, it is easy to see
that the parity in the number of excitations $\mathbf{M}$ is
conserved. This can be proved by using the unitary transformation $
\mathbf{U}(\theta):=\exp\left(i\,\theta\,\mathbf{M}\right)$.
To this end one writes the Hamiltonian~(\ref{eq01}) in the form
$\mathbf{H}=\mathbf{H}_{\hbox{\tiny{RWA}}}+\mathbf{H}_{R}$
where
   \begin{eqnarray}
      \mathbf{H}_{R}&:=&-\frac{1}{\sqrt{N_{a}}}
      \,\sum_{i<j}^{3}\,\mu_{ij}\,\left(\,
      \mathbf{A}_{ij}\,a+\mathbf{A}_{ji}\,a^{\dagger}
      \right)\ .
   \end{eqnarray}
Its transformation under $\mathbf{U}(\theta)$ is
[cf.~\ref{appA}]
   \begin{equation}
      \mathbf{U}(\theta)\,\mathbf{H}\,\mathbf{U}^{\dagger}(\theta)
      =\mathbf{H}_{\hbox{\tiny{RWA}}}+\cos(2\theta)\,\mathbf{H}_{R}
      +\frac{i}{2}\,\sin(2\theta)\,\Big[\,\mathbf{M},\,\mathbf{H}_{R}\,\Big]\ .
   \end{equation}
Then $\exp(i\,\theta\,\mathbf{M})$ is only a symmetry operator when
$\theta=\pi$. Therefore the solutions may only have an even or odd parity
in the total number of excitations $\mathbf{M}$. Thus it is convenient
to define a linear combination of coherent states to preserve the
symmetry of the Hamiltonian:
	\begin{equation}
		\vert\alpha;\,\gamma\}_{\pm}:=\left(\mathbf{1}
		\pm\exp(i\,\pi\,\mathbf{M})\right)
		\,\vert\alpha;\,\gamma\}\ ,
	\end{equation}
which will be called {\it symmetry-adapted coherent states} (SACS) of the
Hamiltonian.

The unnormalised SACS states are given by
   \begin{eqnarray*}
   	\fl
        \vert\alpha;\,\gamma\}&=&\sum_{\nu=0}^{\infty}\,
        \sum_{n_{1},\,n_{2},\,n_{3}\atop n_{1}+n_{2}+n_{3}=N_{a}}\,
        \sqrt{\frac{N_{a}!}{\nu!\,n_{1}!\,n_{2}!\,n_{3}!}}\,\alpha^{\nu}
        \,\gamma_{2}^{n_{2}}\,\gamma_{3}^{n_{3}}
        \,\vert\,\nu;\,n_{1},\,n_{2},\,n_{3}\rangle\\
        \fl
        &=&\sum_{m=0}^{\infty}\ \sum_{n_{2},\,n_{3}=0}^{N_{a}}\,\sqrt{
        \frac{N_{a}!}{(m-\lambda_{2}\,n_{2}-\lambda_{3}\,n_{3})!
        \,(N_{a}-n_{2}-n_{3})!\,n_{2}!\,n_{3}!}}\\
        \fl
        &&\qquad\times\alpha^{m-\lambda_{2}\,n_{2}-\lambda_{3}\,n_{3}}
        \,\gamma_{2}^{n_{2}}\,\gamma_{3}^{n_{3}}
        \,\vert\,m-\lambda_{2}\,n_{2}-\lambda_{3}\,n_{3}\,;
        \,N_{a}-n_{2}-n_{3},\,n_{2},\,n_{3}\rangle\ ,
   \end{eqnarray*}
where in the last expression we have replaced the eigenvalue of the
number of photons $\nu$ by the eigenvalue of the total excitation
number $m=\nu+\lambda_{2}\,n_{2}+\lambda_{3}\,n_{3}$. 
If we now make in the previous expression the substitutions
$\alpha\rightarrow-\alpha$ and
$\gamma\rightarrow\tilde{\gamma}=(\gamma_{1},
\,(-1)^{\lambda_{2}}\gamma_{2},\,(-1)^{\lambda_{3}}\gamma_{3})$,
we can show that in the superposition
$\vert\alpha;\,\gamma\}\pm\vert-\alpha;\,\tilde{\gamma}\}$ either the
odd or the even parity contributions of the total number of excitations
cancel, but not both, leaving a state with $\mathbf{M}$-parity well
defined, i.e.,
   \begin{eqnarray}
      \fl
      &&\vert\alpha;\,\gamma\}_{\pm}:=\vert\alpha;\,\gamma\}
      \pm\vert-\alpha;\,\tilde{\gamma}\}\nonumber\\
      \fl
      &&\quad=
      \sum_{m=0}^{\infty}(1\pm(-1)^{m})\,\sum_{n_{2},\,n_{3}=0}^{N_{a}}\,\sqrt{\frac{N_{a}!}{
      (m-\lambda_{2}\,n_{2}-\lambda_{3}\,n_{3})!
      \,(N_{a}-n_{2}-n_{3})!\,n_{2}!\,n_{3}!}}\nonumber\\
      \fl
      &&\quad\times(\alpha)^{m-\lambda_{2}\,n_{2}-\lambda_{3}\,n_{3}}
      \,\gamma_{2}^{n_{2}}\,\gamma_{3}^{n_{3}}
      \,\vert\,m-\lambda_{2}\,n_{2}-\lambda_{3}\,n_{3}\,;\,N_{a}-n_{2}-n_{3},\,n_{2},\,n_{3}\rangle\ .
      \label{eq038}
   \end{eqnarray}
Thus $\vert\alpha;\,\gamma\}_{+}$ contains only terms with even values
of $m$, while $\vert\alpha;\,\gamma\}_{-}$ has only terms with odd
values of $m$. For this reason these states are orthogonal.

The reproducing kernel of these new states takes the form
	\begin{eqnarray}
		&&{}_{\pm}\{\alpha;\,\gamma_{2},\,\gamma_{3}
		\vert\alpha^{\prime};\,\gamma^{\prime}_{2},
		\,\gamma^{\prime}_{3}\}_{\pm}\nonumber\\
		&&=2\,\Big[\exp(\alpha^{\ast}\alpha^{\prime})\,
		(\gamma^{\ast}\cdot\gamma^{\prime})^{N_{a}}
		\pm\exp(-\alpha^{\ast}\alpha^{\prime})\,
		(\gamma^{\ast}\cdot\tilde{\gamma}^{\prime})^{N_{a}}
		\Big]\ .
	\end{eqnarray}

The energy surface for the SACS is given by the expression
   \begin{eqnarray}
   	\fl
      &&{}_{\pm}\{\alpha,\,\gamma\vert\,\mathbf{H}\,\vert\alpha,\,\gamma\}_{\pm}
      \nonumber\\
      \fl
      &&=2\,\Omega\,|\alpha|^{2}\,\Big[\exp(|\alpha|^{2})\,(\gamma^{\ast}\cdot\gamma)^{N_{a}}
      \mp\exp(-|\alpha|^{2})\,(\gamma^{\ast}\cdot\tilde{\gamma})^{N_{a}}\Big]\nonumber\\
      \fl
      &&+2\,N_{a}\,\sum_{i=1}^{3}\omega_{i}\,|\gamma_{i}|^{2}\,\Big[
      \exp(|\alpha|^{2})\,(\gamma^{\ast}\cdot\gamma)^{N_{a}-1}
      \pm(-1)^{\lambda_{i}}\,\exp(-|\alpha|^{2})
      \,(\gamma^{\ast}\cdot\tilde{\gamma})^{N_{a}-1}\Big]\nonumber\\
      \fl
      &&+\sqrt{N_{a}}\,(\alpha+\alpha^{\ast})\,\sum_{i<j=1}^{3}\,\mu_{ij}
      \,(1-(-1)^{\lambda_{i}+\lambda_{j}})\nonumber\\
      \fl
      &&\quad\times\Big[\exp(|\alpha|^{2})\,(\gamma_{i}^{\ast}\,\gamma_{j}
      +\gamma_{j}^{\ast}\,\gamma_{i})\,(\gamma^{\ast}\cdot\gamma)^{N_{a}-1}
      \nonumber\\
      \fl
      &&\qquad\pm\exp(-|\alpha|^{2})\,((-1)^{\lambda_{i}}\,\gamma_{i}^{\ast}\,\gamma_{j}
      +(-1)^{\lambda_{j}}\,\gamma_{j}^{\ast}\,\gamma_{i})
      \,(\gamma^{\ast}\cdot\tilde{\gamma})^{N_{a}-1}\Big]\ .
   \end{eqnarray}
Then, substituting the polar form of the complex variables $\alpha$,
$\gamma_k$, with $k=1,2,3$ as before, and using the minima presented
in section 4, we obtain the energy surface in the normal and
superradiant regimes for the $V$-configuration under the double
resonance condition.

In \ref{appB} we calculate the expectation values of
matter, field, and matter-field operators with respect to the SACS

\section{Comparison of Coherent States and SACS Approximations}

The approximation to the ground state with the SACS is obtained by
minimising the expectation value of the Hamiltonian with respect to
these states. In this form we obtain approximations to the ground
state (even SACS), and to the first-excited state (odd SCAS). However,
the minimisation is much more complicated than in the ordinary
coherent state case. Nevertheless, and based on the results for
two-level systems~\cite{nahmad2012}, we can make the assumption that
the value that we will obtain will be a very small correction to the
one obtained using the ordinary coherent states.  Indeed this appears
to be the case except in a very small vicinity of the points around
the place were the quantum phase transition takes place, i.e., around
the curve which determines the separation between the normal and
collective regimes. This difference of course goes to zero in the
thermodynamic limit.

In this contribution we will study only the $V$-configuration in
double-resonance because we have analytic expressions for the minimum
points [cf. Eqs.~(\ref{eq025}-\ref{eq027})]. We will take the values
$\Omega=1$, $\omega_{1}=0$, and $\omega_{2}=\omega_{3}=1$ in all the
calculations presented in this work.

Because of the symmetry appearing in Eqs.~(\ref{eq026}-\ref{eq027}),
it is convenient to introduce the parametrization of the dipole
interaction parameters as
	\begin{equation}
		\mu_{12}=\mu\,\cos\theta\ ,\quad\mu_{13}=\mu\,\sin\theta\ .
	\end{equation}

Thus by substituting $\varrho_{c}$, $\varrho_{2\,c}$ and
$\varrho_{3\,c}$ as given in Eqs.~(\ref{eq025}-\ref{eq027}), the
energy per atom in the superradiant region $\mu^{2}>1/4$ is given by
	\begin{eqnarray}
		E_{\pm}/N_{a}&=&E_{coh}/N_{a}\pm\frac{2\,\left(\mu^{2}
                -\frac{1}{16\,\mu^{2}}\right)}{1\pm\left(
                2\mu\,\exp(\mu^{2}-\mu^{-2}/16)\right)^{2N_{a}}}\ ,
       \end{eqnarray}
while for the corresponding case in the coherent state approximation
       \begin{eqnarray}
		E_{coh}/N_{a}&=&-\left(\mu-\frac{1}{4\,\mu}\right)^{2}\ .
	\end{eqnarray}

It is immediate that in the limit when $N_{a}\rightarrow\infty$ or
$\mu\rightarrow\infty$ both expressions are identical. For the normal
regime one has $E_+/N_a=0$, $E_-/N_a= 1/(2 N_a)$ and $E_{coh}/N_a=0$.

\begin{figure}[h]
\begin{center}
    \includegraphics[scale=0.3]{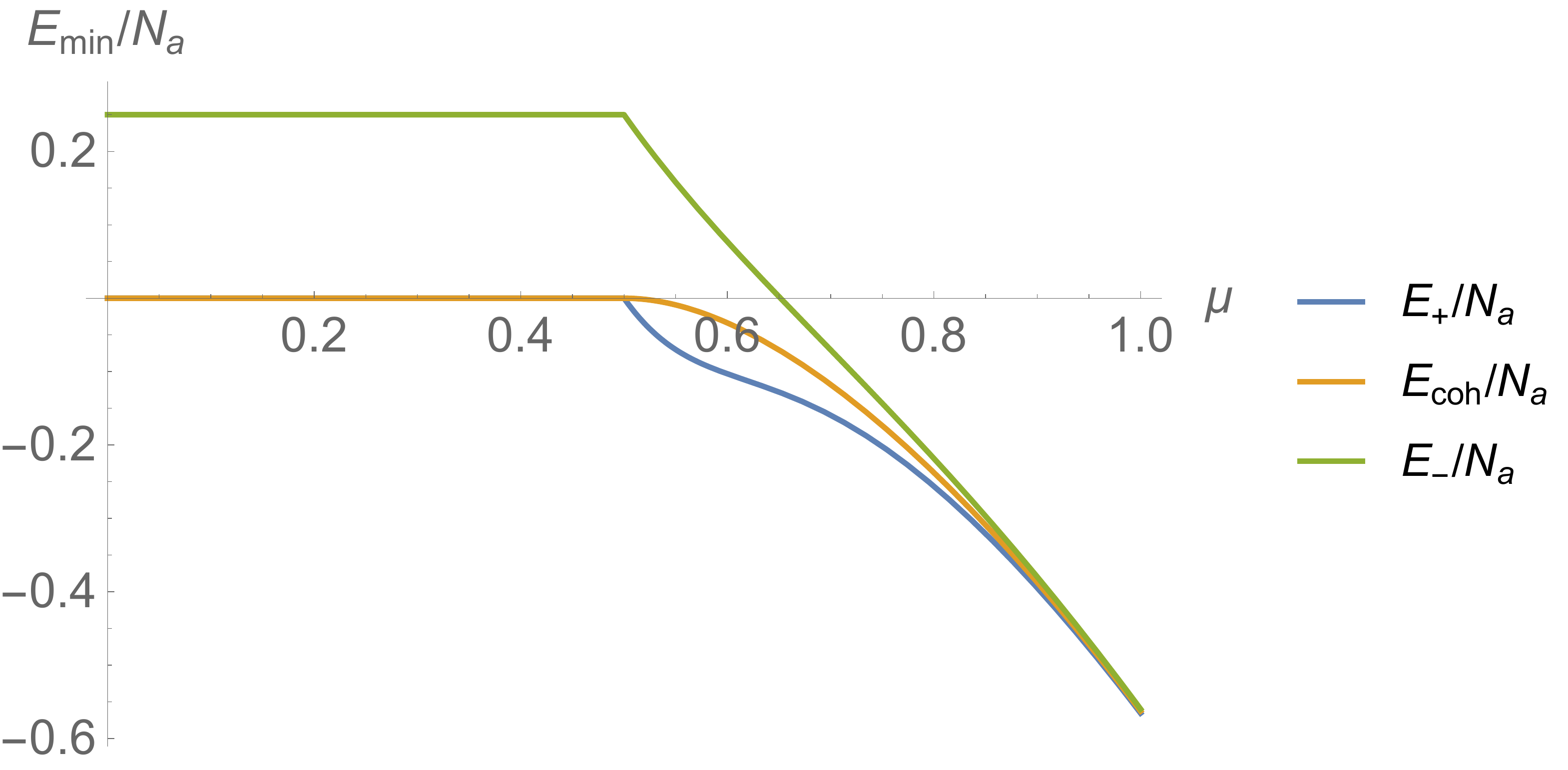}
\caption{%
(Colour online) Ground state energy comparison for the coherent state
(orange) and SACS approximations. We consider $N_{a}=2$ atoms,
$\Omega=\omega_{2}=\omega_{3}=1$, and $\omega_{1}=0$. We use the
variable $\mu$ to denote the intensity of the dipolar
interactions. The expectation value for the even and odd SACS states
are the lowest (blue) and the highest (green) curves, respectively.
\label{fig01}
}
\end{center}
\end{figure}

In Fig.~\ref{fig01} we show a plot of the energy for the coherent
state and SACS approximations to the ground and first excited state
energy for $N_{a}=2$ atoms. We observe that the even SACS state gives
an energy below that of the coherent state approximation. The odd SACS
state corresponds to the estimation of the energy of the first-excited
state. We note that all the SACS estimations approach the coherent one
as the intensity of the interaction grows. When the number of
particles is larger, the form of the SACS energies approximates the
coherent state result; even for $N_{a}=10$ they are very difficult to
distinguish.

\begin{figure}[h]
\begin{center}
    \includegraphics[scale=0.19]{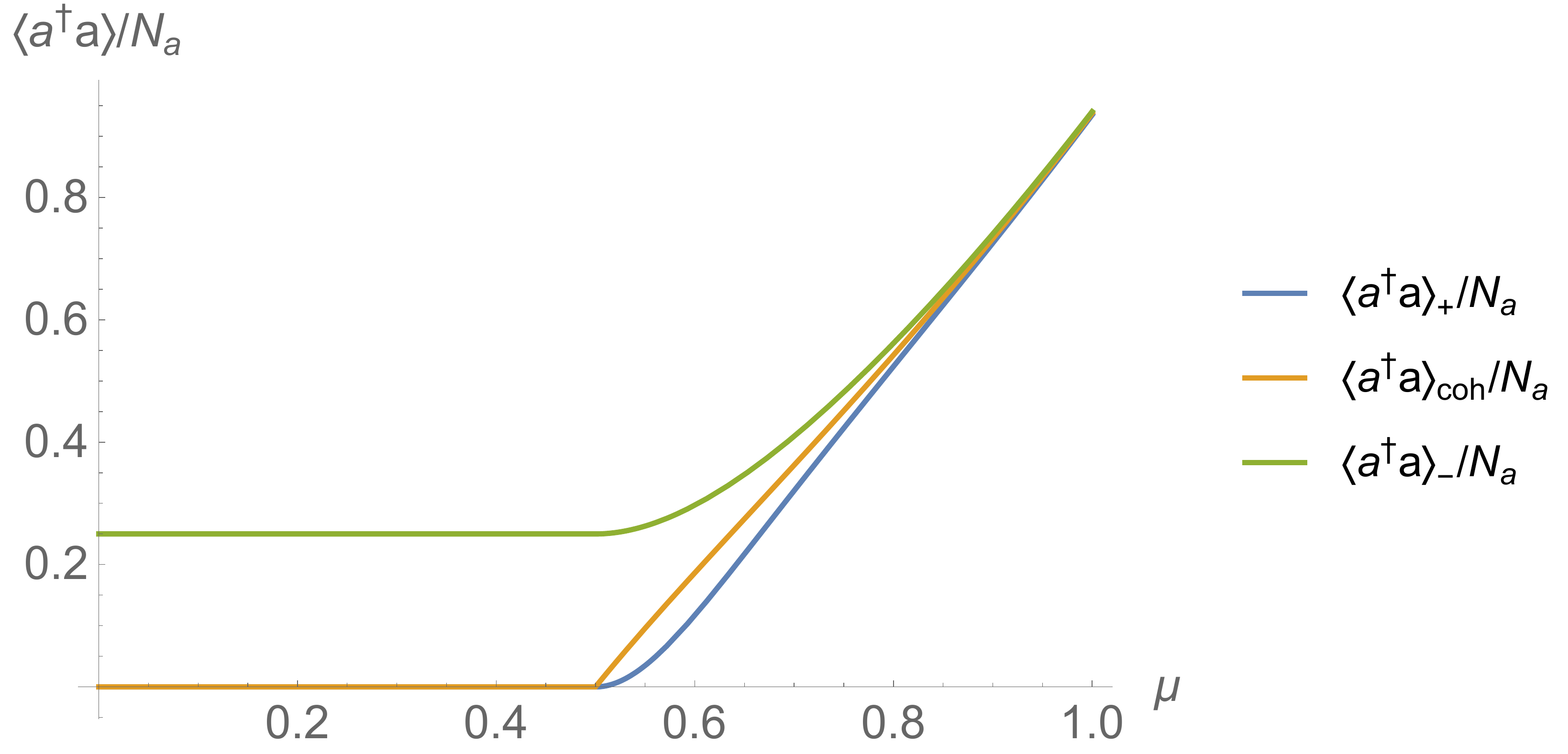}
    \includegraphics[scale=0.19]{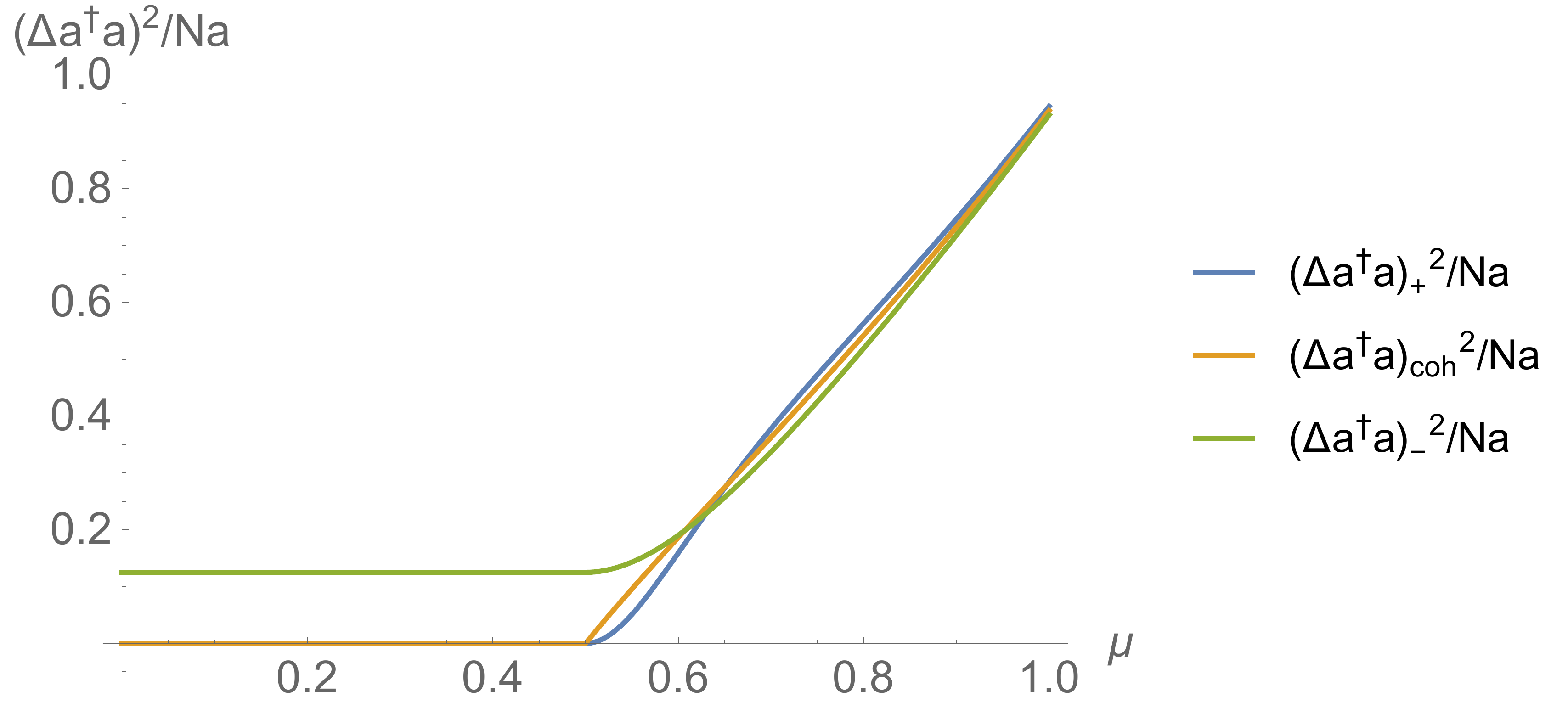}
\caption{%
(Colour online) Expectation value (left) and squared fluctuation
(right) of the number of photons per atom, for the coherent state and
SACS approximations, with $N_{a}=2$ atoms. $\mu$ denotes the intensity
of the dipolar interactions.
\label{fig02}
}
\end{center}
\end{figure}

We can calculate also the expectation values of other observables of
the system. In Fig.~\ref{fig02} (left) we show a plot of the
expectation value of the number of photons per atom vs. the intensity
of the dipole interaction for the coherent state and SACS state
approximations to the ground state, for $N_{a}=2$ atoms. As before,
they all converge when $N_{a}$ grows. In Fig.~\ref{fig02} (right) we
also show the behaviour of the squared fluctuation in the number of
photons per atom.

\begin{figure}[h]
\begin{center}
    \includegraphics[scale=0.18]{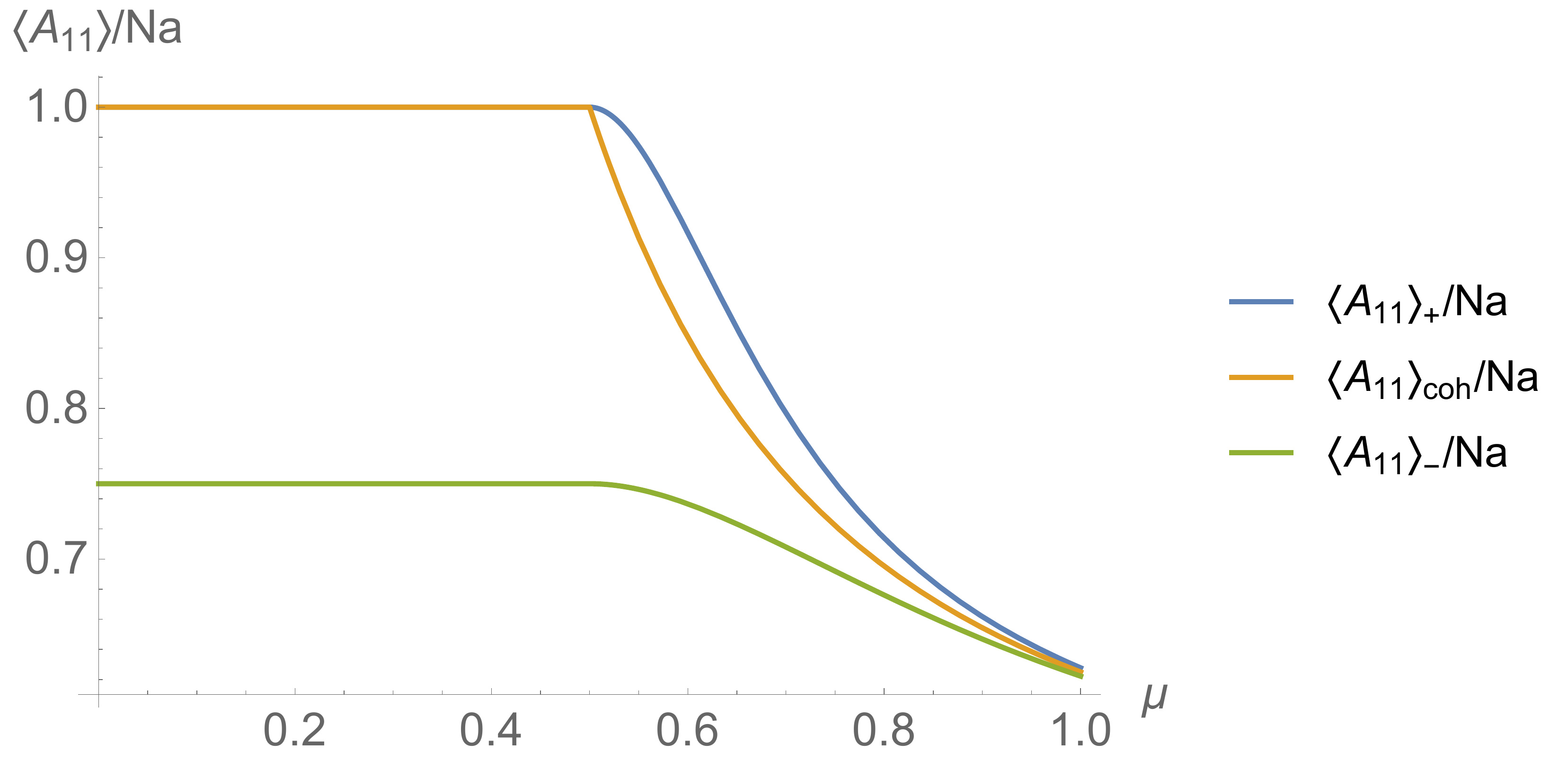} \quad
    \includegraphics[scale=0.18]{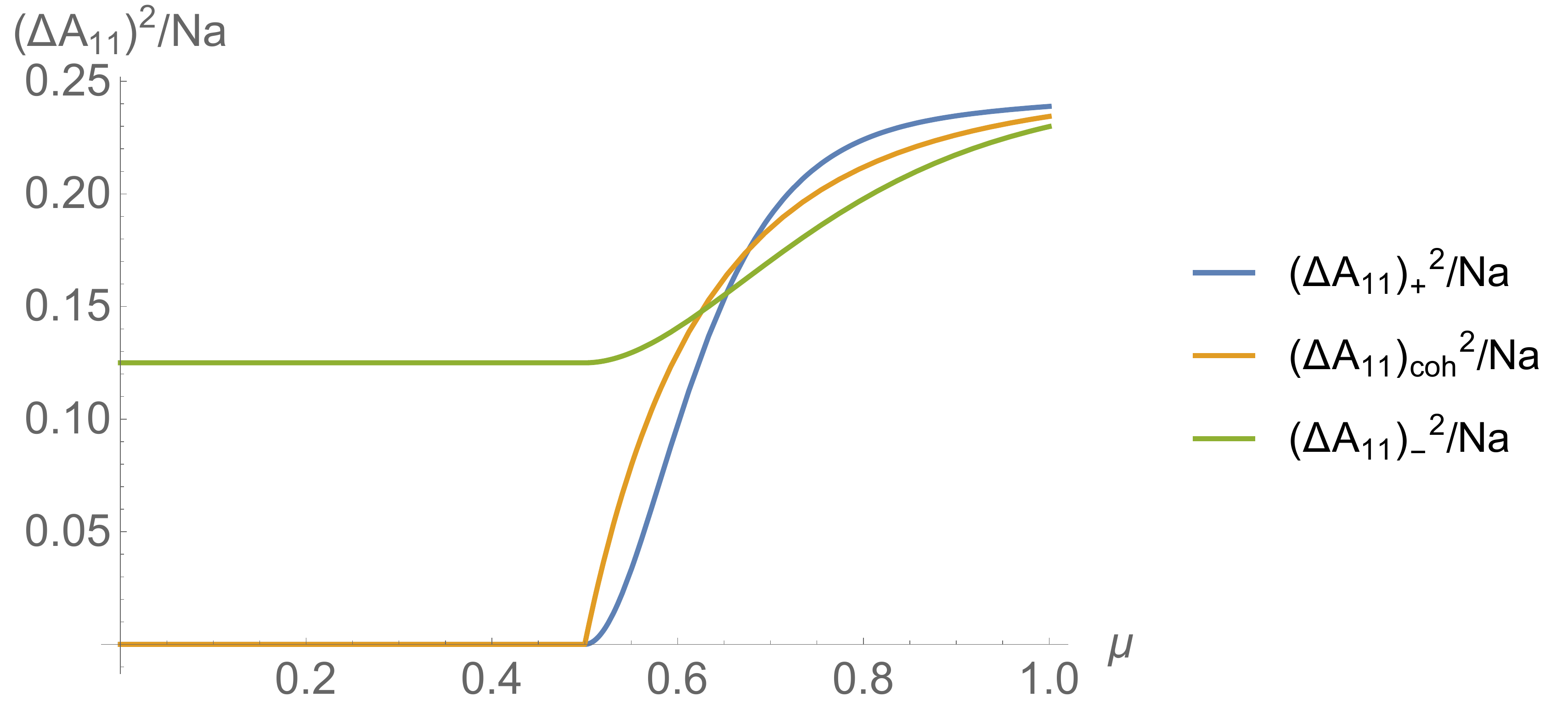}
\caption{%
(Colour online) Comparison of the expectation value of the number
atoms in the lowest energy level $i=1$ per atom for the coherent state
and SACS approximations (left) and squared fluctuation of the atomic
population in energy level $i=1$ per atom for the coherent state and
SACS approximations. We considered $N_{a}=2$ atoms, and
$\mu$ denotes the intensity of the dipolar interactions.
\label{fig03}
}
\end{center}
\end{figure}

\begin{figure}[h]
\begin{center}
    \includegraphics[scale=0.18]{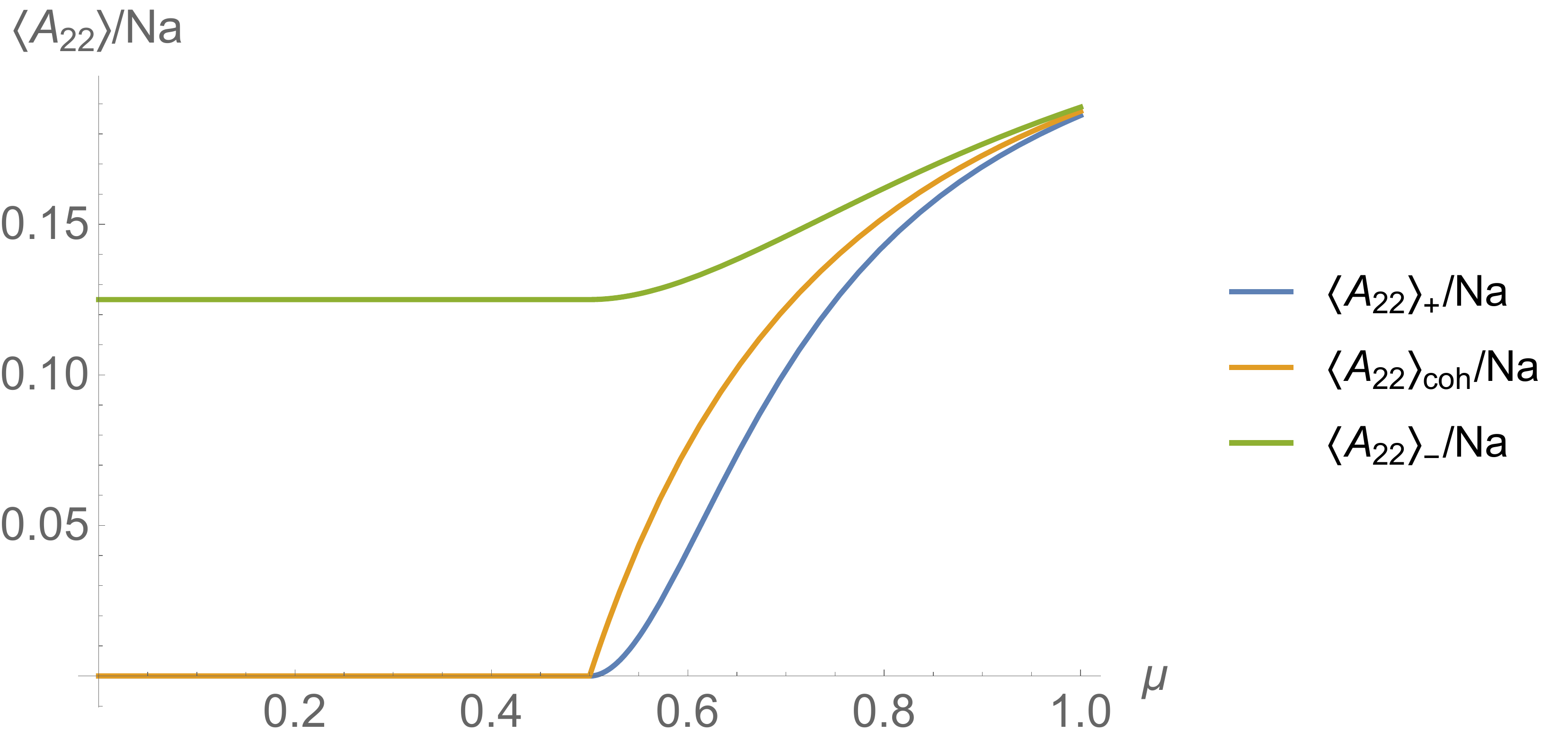} \quad
    \includegraphics[scale=0.18]{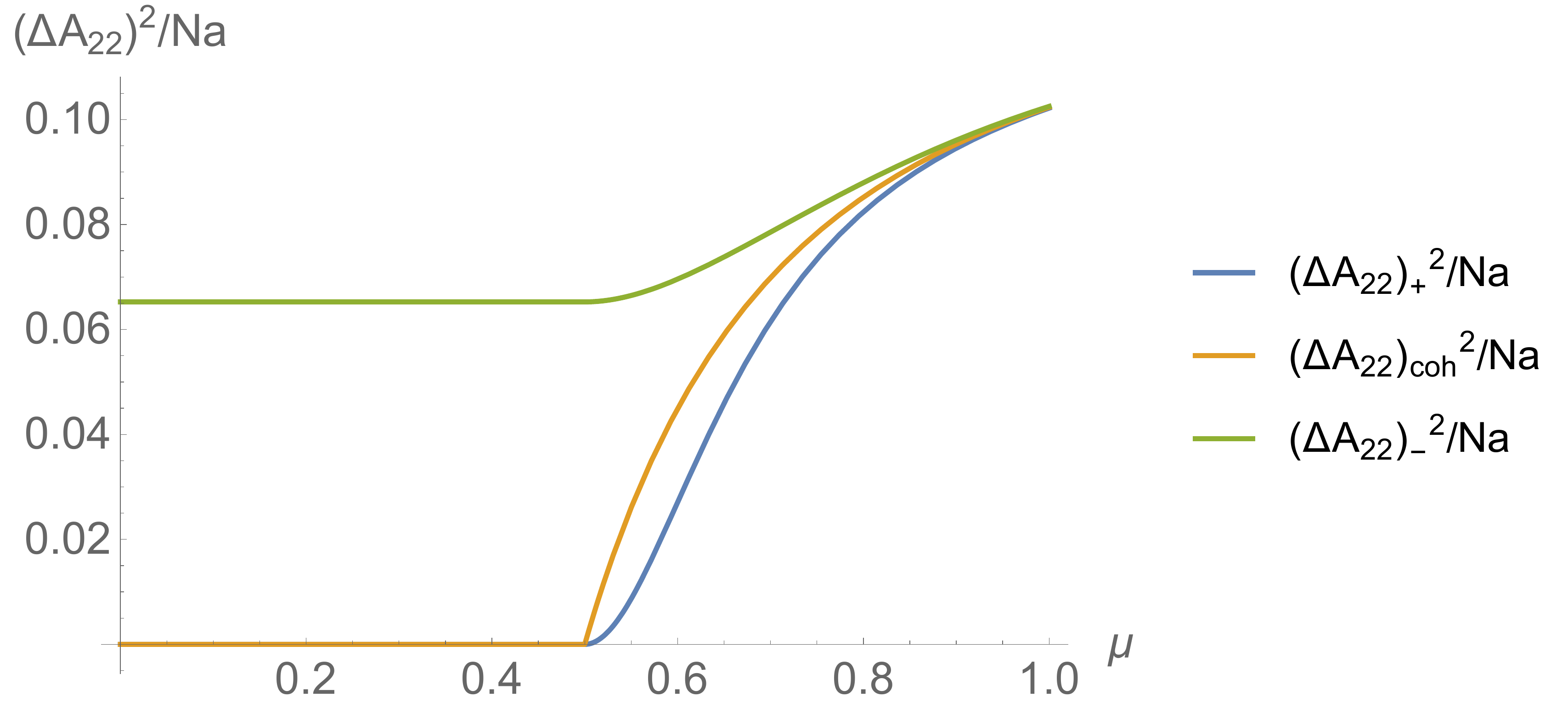}
\caption{%
(Colour online) Comparison of the expectation value of the number
atoms in the middle energy level $i=2$ per atom (left), and their
corresponding squared fluctuations (right), for the coherent state and
SACS approximations. We considered $N_{a}=2$ atoms, and
$\mu$ denotes the intensity of the dipolar interactions.
\label{fig04}
}
\end{center}
\end{figure}

In Figs.~\ref{fig03}-\ref{fig04} we present the expectation values for
the number of atoms in level $i$, with $i=1,2$, normalised by the
number of atoms, and their corresponding fluctuations, respectively,
using $N_{a}=2$ atoms.  The expectation value for the odd SACS is the
lowest of the curves. This behaviour is to be expected because for the
odd SACS we have contribution of atoms in the excited states, which
diminishes the number of atoms in the lowest level.  Exactly the
opposite happens for the expectation values of the occupation of the
excited levels, as the left side of figure~\ref{fig04} shows. In the
right side of figure~\ref{fig03} we show the squared fluctuation for
the occupation of the lowest atomic level. The squared fluctuation for
the coherent and even SACS states is zero in the normal regime
$\mu\le0.5$, as is to be expected; then in the collective regime they
increase and separate, with the value for the coherent state growing
faster. In the normal region the value of the squared fluctuation for
the odd SACS is greater, as it also should be. For larger $\mu$ the
coherent state fluctuation seems to be the average of the SACS
ones. When the intensity of the interaction increases the curves tend
to the same value. The behaviour for the fluctuation of the occupancy
of the other atomic levels is similar. The squared fluctuation per
atom for the coherent and even SACS states is shown for level $i=2$ in
Fig.~\ref{fig04} (right). We see that it is zero in the normal regime
$\mu\le0.5$, for the even SACS and the coherent states. In the normal
region the value of the squared fluctuation per atom for the odd SACS
state is again greater.  The asympotic behaviour is the same as for
$i=1$. Because the behaviour for the fluctuation of the occupancy
levels $i=2,\,3$ is similar, the latter is not shown.

In fact there is a simple relation between the expectation values for
levels $i=2,\,3$ and the expectation value of the lowest level $i=1$:
   \begin{equation}
   \fl 
      \langle\mathbf{A}_{22}\rangle=N_{a}\,\left(1
      -\frac{1}{N_{a}}\langle\mathbf{A}_{11}\rangle\right)\,\cos^{2}\theta\ ,\quad
      \langle\mathbf{A}_{33}\rangle=N_{a}\,\left(1
      -\frac{1}{N_{a}}\langle\mathbf{A}_{11}\rangle\right)\,\sin^{2}\theta\ .
   \end{equation}

\begin{figure}[h]
\begin{center}
    \includegraphics[scale=0.18]{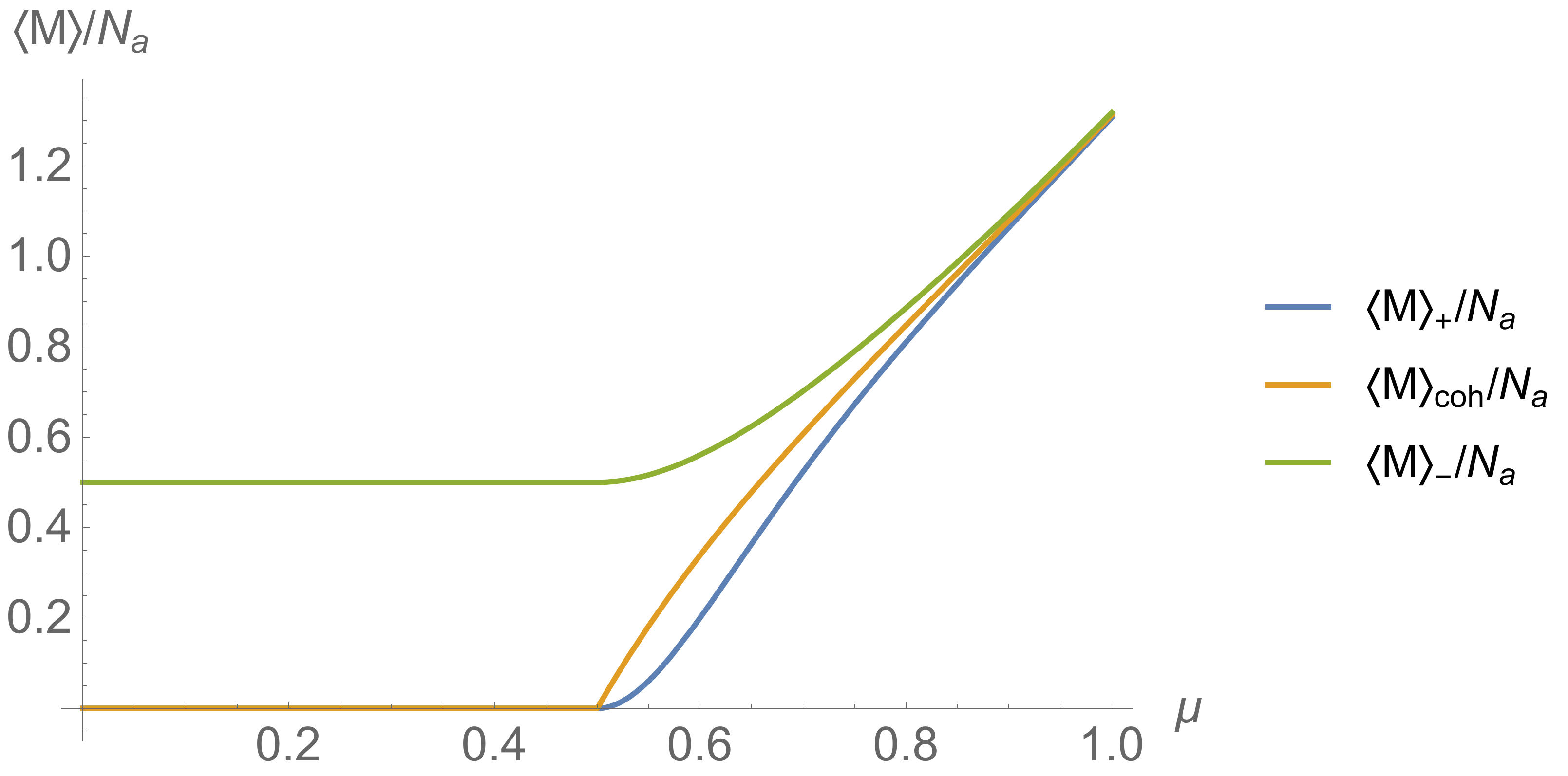} \quad
    \includegraphics[scale=0.18]{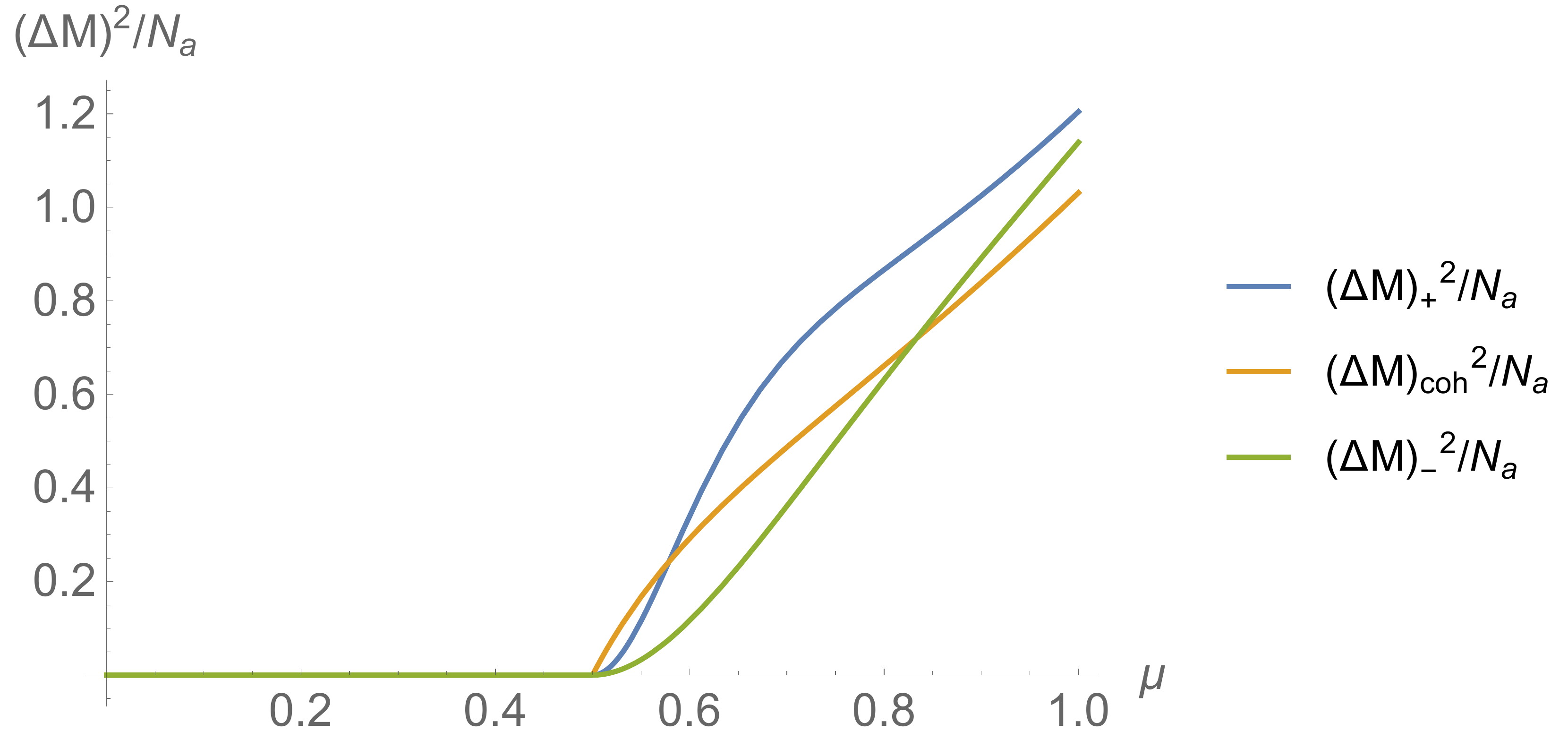}
\caption{%
(Colour online) Expectation value of the number of excitations
$\mathbf{M}$ (left) and its corresponding squared fluctuation
(right). We considered $N_{a}=2$ atoms, and
$\Omega=\omega_{2}=\omega_{3}=1$ and $\omega_{1}=0$. The expectation
value per atom for the coherent and even SACS states is zero in the
normal regime, $\mu\le0.5$. In the collective regime they increase and
separate, with the value for the coherent state growing faster.
\label{fig05}
}
\end{center}
\end{figure}

The expectation value of the number of excitations $\mathbf{M}$ for
our approximations to the ground and first excited states are shown in
Fig.~\ref{fig05} (left). The behaviour is similar to the one exhibited
for the expectation value of the number of photons or the occupancy of
the upper atomic levels. The expectation value per atom for the
coherent and even SACS states is zero in the normal regime,
$\mu\le0.5$, then in the collective regime they increase and separate
with the value for the coherent state growing faster. When the
intensity of the interaction increases all the curves tend to the same
value. Fig.~\ref{fig05} (right) shows the value of the fluctuations in
the total number of excitations for the different approximations.  While the
even and coherent estimations are the same for the normal regime, we
observe that their values are very different in the collective
regime.

\begin{figure}[h]
\begin{center}
    \includegraphics[scale=0.3]{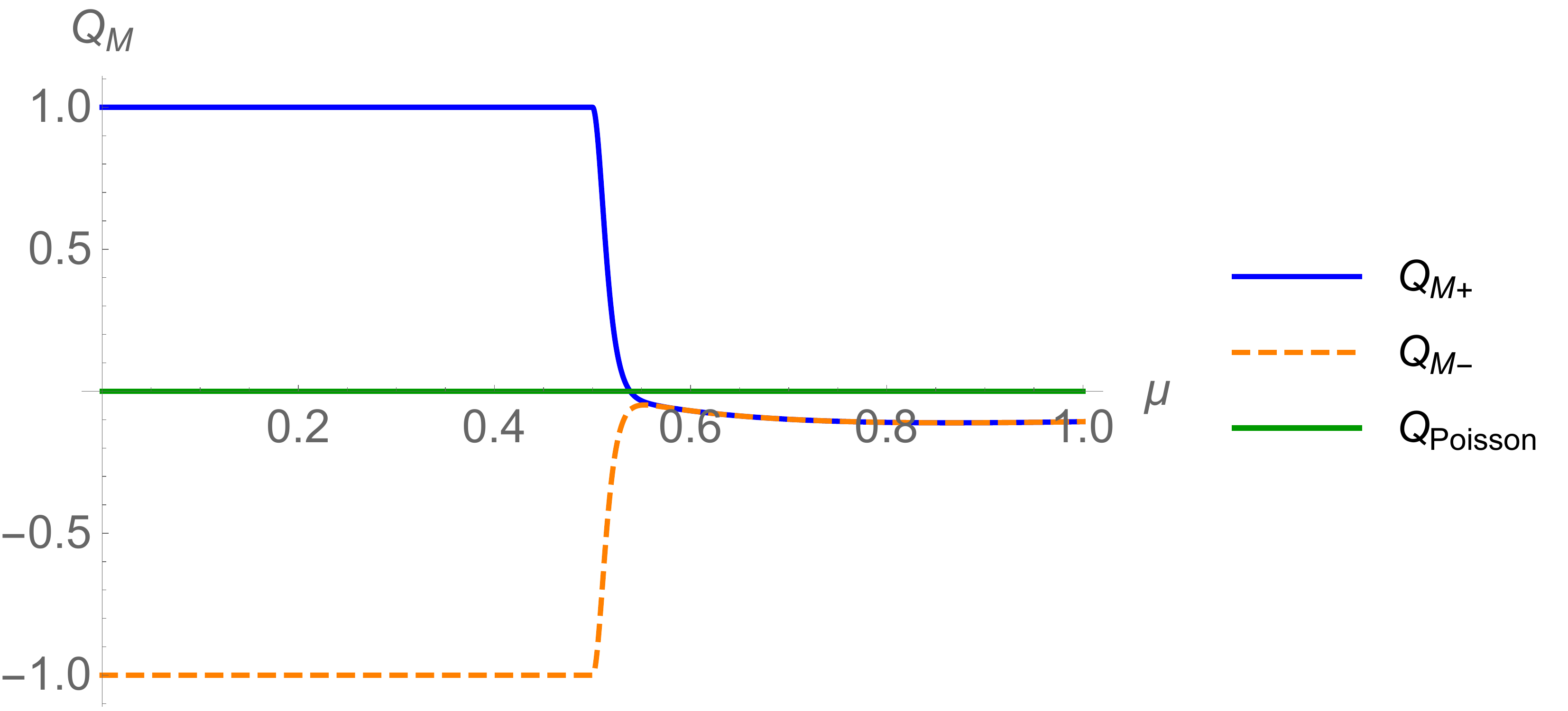}
\caption{%
Analogous $Q$ Mandel parameter for the number of excitations
$\mathbf{M}$, with $N_{a}=2$ atoms. We observe that in the normal
regime the behaviour of the three approximations is very
different. For the even SACS state approximation we have the value 1,
where we can say that the behaviour is superpoissonian, and it starts
to decrease at the phase transition ($\mu=0.5$) until it crosses to
negative values at $\mu\approx0.54$, becoming subpoissonian. The odd
SACS state approximation starts at -1 for the normal regime and starts
to grow at the critical point, i.e., it is subpoissonian, and then
grows until it reaches the value of the even SACS approximation at
$\mu\approx0.56$. After this point they tend asymptotically to zero
from negative values of $Q$.
\label{fig06}
}
\end{center}
\end{figure}

To determine the statistical behaviour of the total number of
excitations $\mathbf{M}$ we define the analogous of the Mandel
parameter as
   \begin{equation}
      Q_{\mathbf{M}}:=\frac{(\Delta\mathbf{M})^{2}}{
      \langle\mathbf{M}\rangle}-1\ .
   \end{equation}
In Fig.~\ref{fig06} we show this Mandel parameter. In the normal
regime the behaviour of the three approximations is very different.
For the even SACS state approximation the value is 1, i.e., the
behaviour is superpoissonian, and it starts to decrease at the phase
transition $\mu=0.5$ until it crosses to negative values at
$\mu\approx0.54$, becoming a subpoissonian distribution. The odd SACS
state approximation has value -1 for the normal regime, i.e., it is
subpoissonian, and starts to grow at the critical point, never
crossing to positive values, reaching the value of the even SACS state
approximation at $\mu\approx0.56$. After this point they tend
asymptotically to zero from negative values of $Q$.

In contrarst, the coherent state approximation gives a poissonian
distribution throughout. The SACS state estimation for the photon
number distribution function is different than for the coherent state
case because in the latter we have components in the states with an
even and odd number of excitations.  Using Eq.~(\ref{eq038}) and
substituting the minima given in section 4, one gets, for the
$V$-configuration in the double-resonance condition
	\begin{equation}
		{\cal P}_{\pm}(\nu)=\left\{
		\begin{array}{cl}
		\frac{\overline{\nu}^{\nu}}{\nu!}\,\frac{(2\mu)^{N_{a}}\pm(-1)^{\nu}(2\mu)^{-N_{a}}}{
		(2\mu)^{N_{a}}\,\exp(\overline{\nu})\pm(2\mu)^{-N_{a}}\,\exp(-\overline{\nu})}
		&,\ \mu^{2}-\frac{1}{4}\ge0\ ;\\%
		[1.2ex]
		\frac{1}{2}(1\pm1)\,\delta_{\nu\,0}+\frac{1}{4}(1\mp1)(\delta_{\nu\,0}+\delta_{\nu\,1})
		&,\ \mu^{2}-\frac{1}{4}\le0\ ,
		\end{array}\right.
	\end{equation}
where we have defined
	\[
		\overline{\nu}:=N_{a}\frac{\left(\mu^{2}-\frac{1}{4}\right)
		\left(\mu^{2}+\frac{1}{4}\right)}{\mu^{2}}\ .
	\]
For comparison, we write the expression for the normal coherent state:
	\begin{equation}
		{\cal P}_{coh}(\nu)=\left\{
		\begin{array}{cl}
		\frac{\overline{\nu}^{\nu}}{\nu!}\,\exp(-\overline{\nu})&,\ \mu^{2}-\frac{1}{4}\ge0\ ;\\%
		[0.8ex]
		\delta_{\nu\,0}&,\ \mu^{2}-\frac{1}{4}\le0\ .
		\end{array}\right.
	\end{equation}

\begin{figure}[ht]
\begin{center}
    \includegraphics[width=0.4\linewidth]{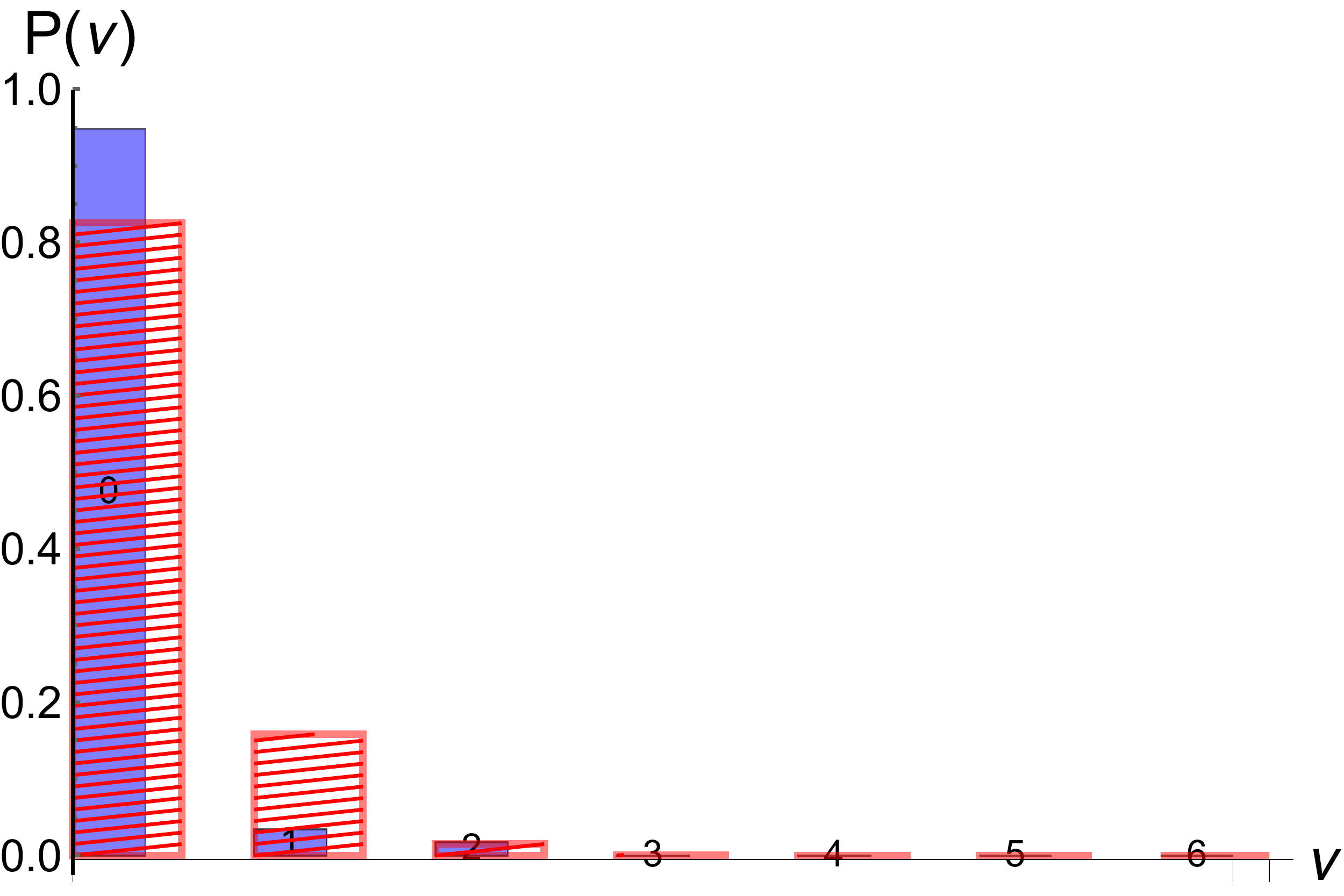}
    \quad \includegraphics[width=0.4\linewidth]{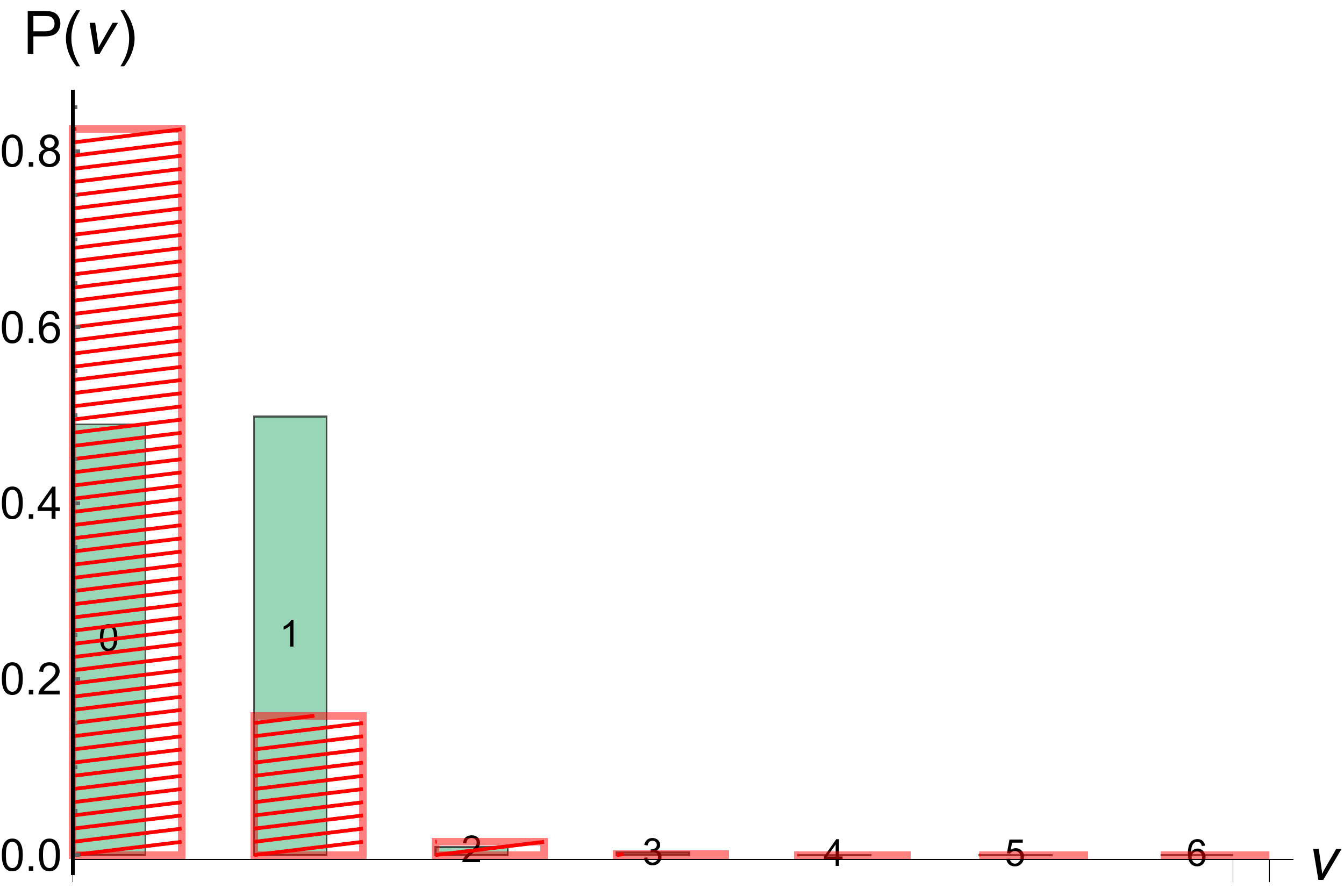}\\[2mm]
    \includegraphics[width=0.4\linewidth]{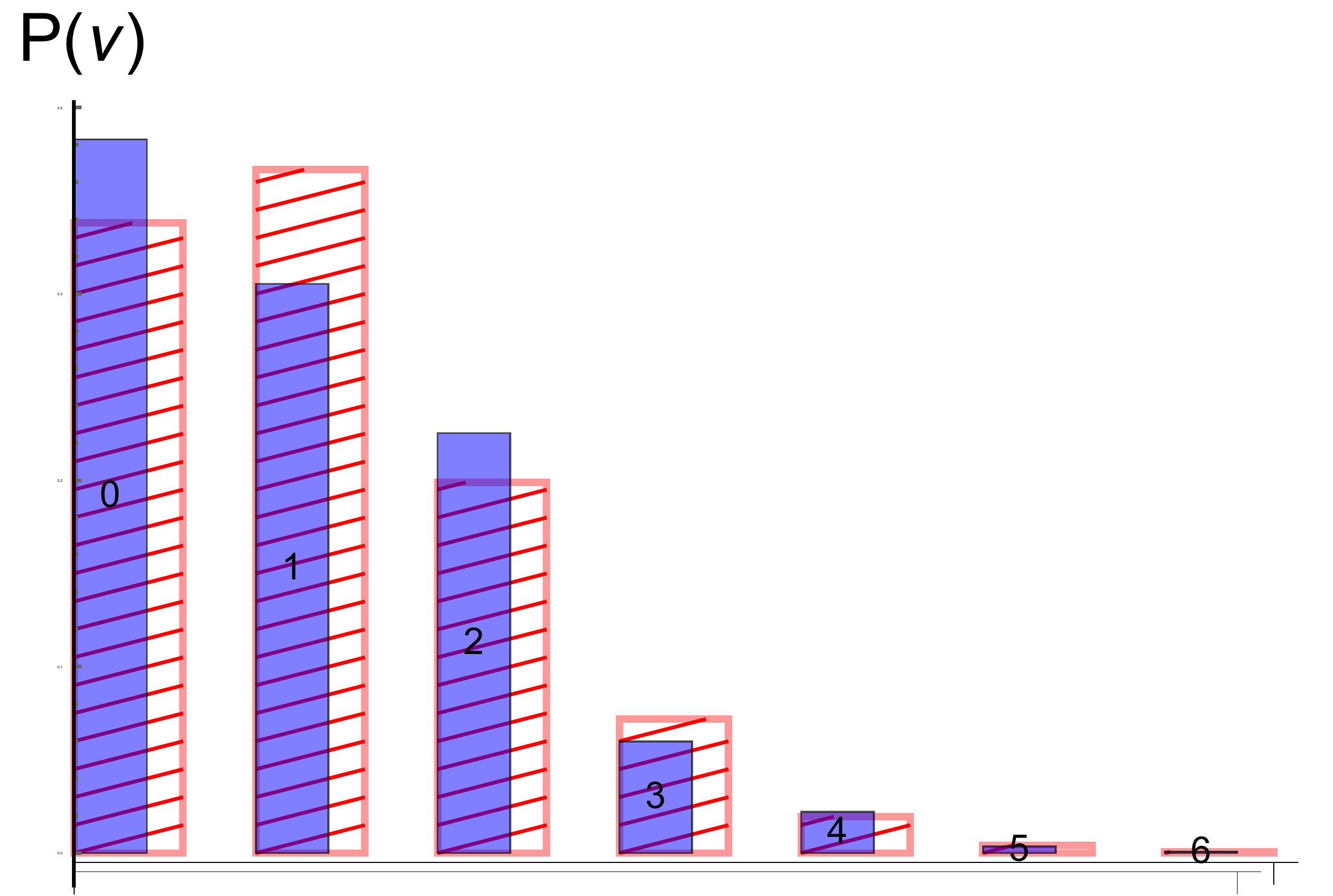}
    \quad \includegraphics[width=0.4\linewidth]{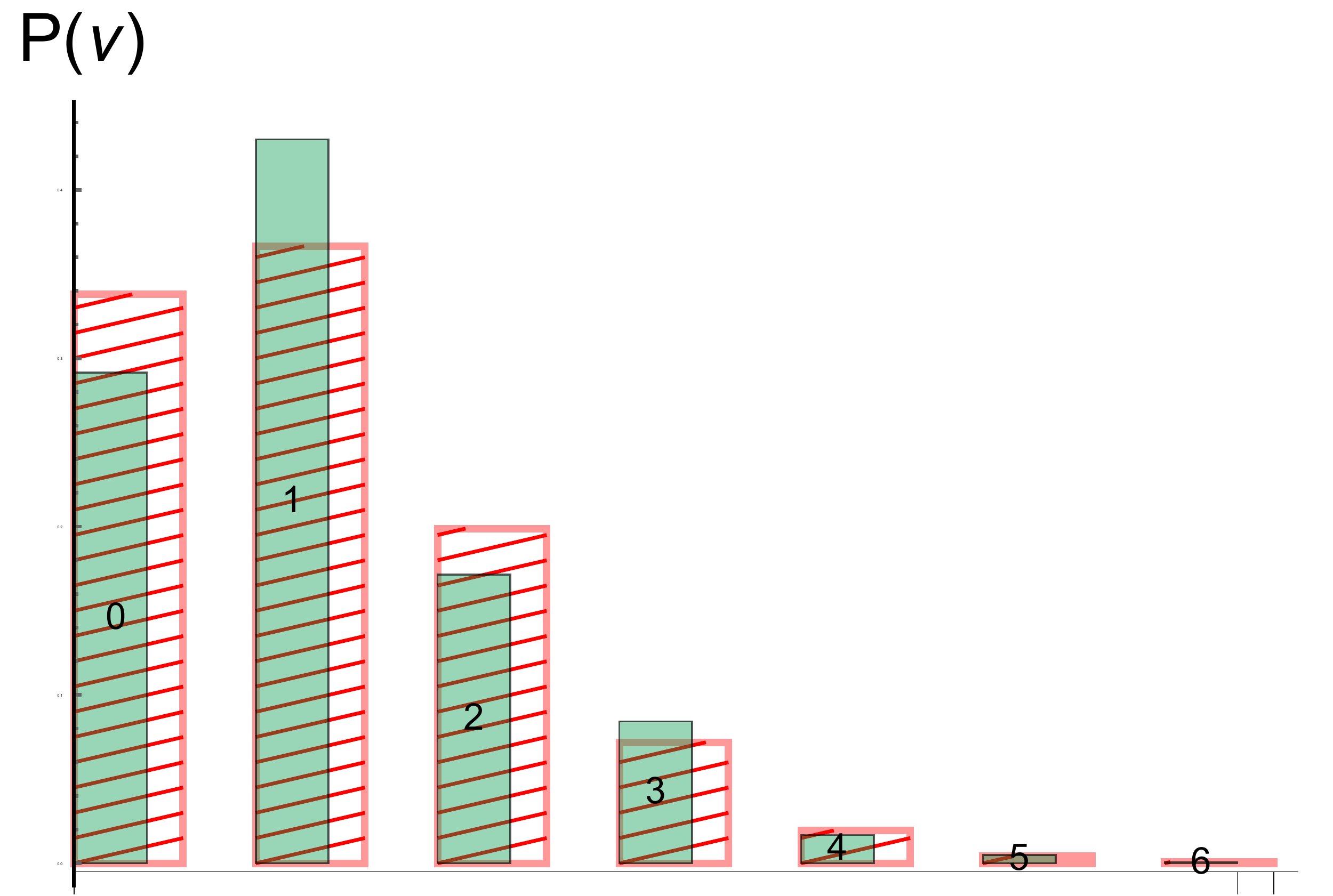}\\[2mm]
    \includegraphics[width=0.4\linewidth]{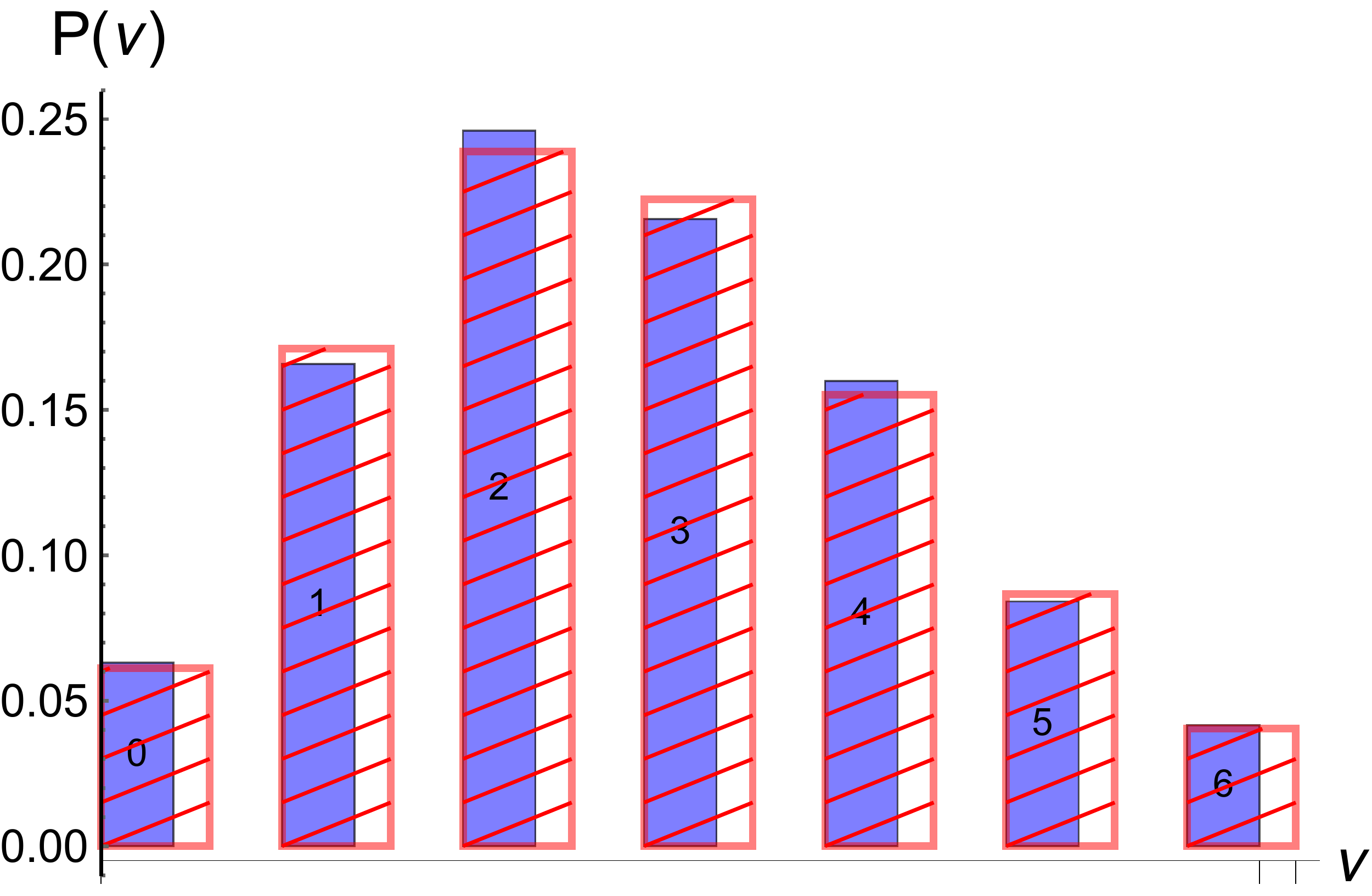}
    \quad \includegraphics[width=0.4\linewidth]{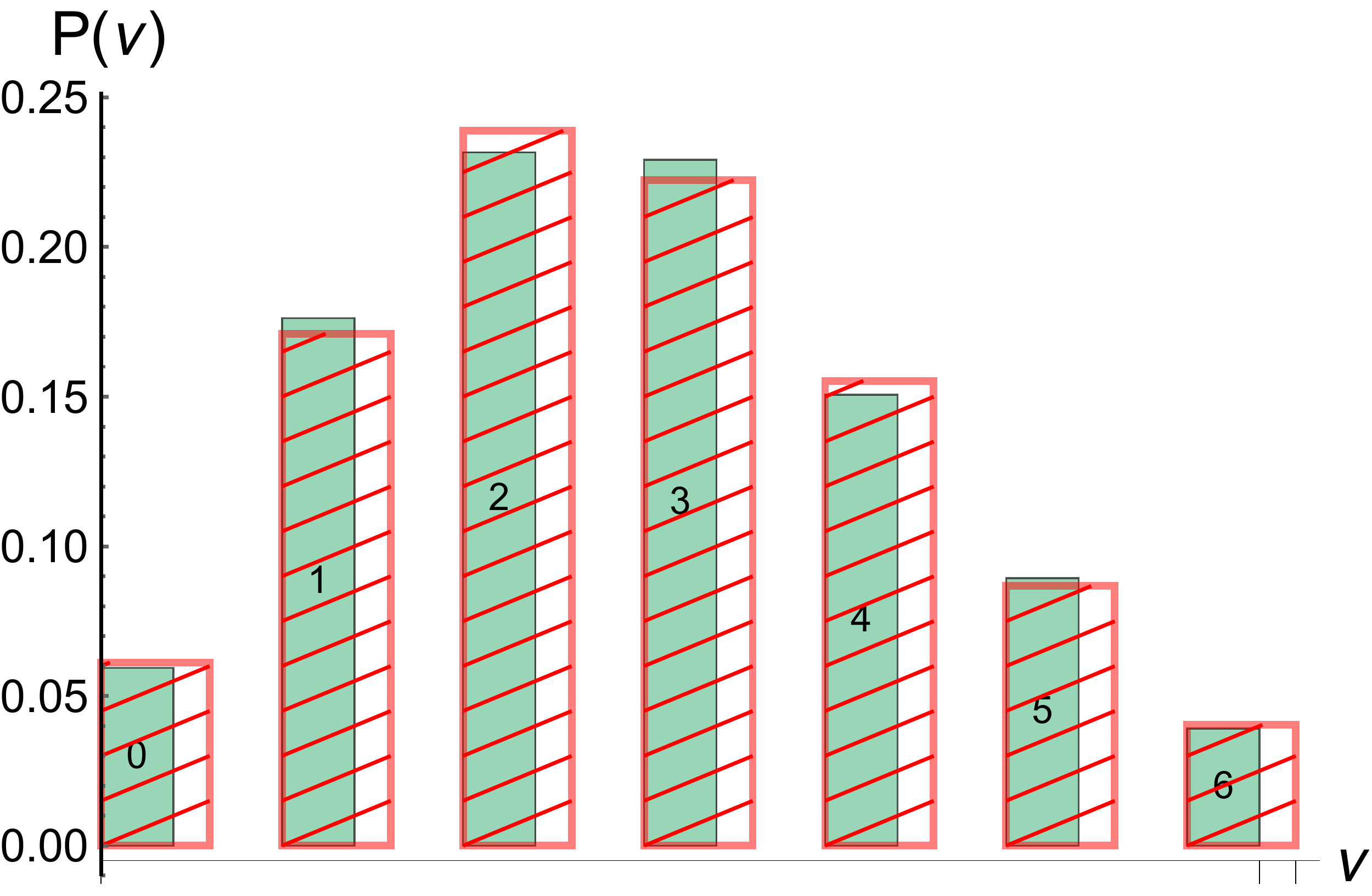}
\caption{%
Comparison between the probability distribution of the number of
photons $\nu$ for the even (left) and odd (right) SACS states (solid
bars) and the coherent state (hatched bars) approximation, in
different regions of the dipole interaction parameter space.  We
considered $N_{a}=2$ atoms. The plots correspond to dipolar
intensities $\mu=0.55,\,0.8,\,1.2$, from top to bottom. We can
appreciate that both distributions differ more as we approach the
phase transition, which occurs at $\mu=0.5$. For greater values of
$\mu$ the distributions are practically indistinguishable.
\label{fig07}
}
\end{center}
\end{figure}

\begin{figure}[ht]
\begin{center}
    \includegraphics[scale=0.3]{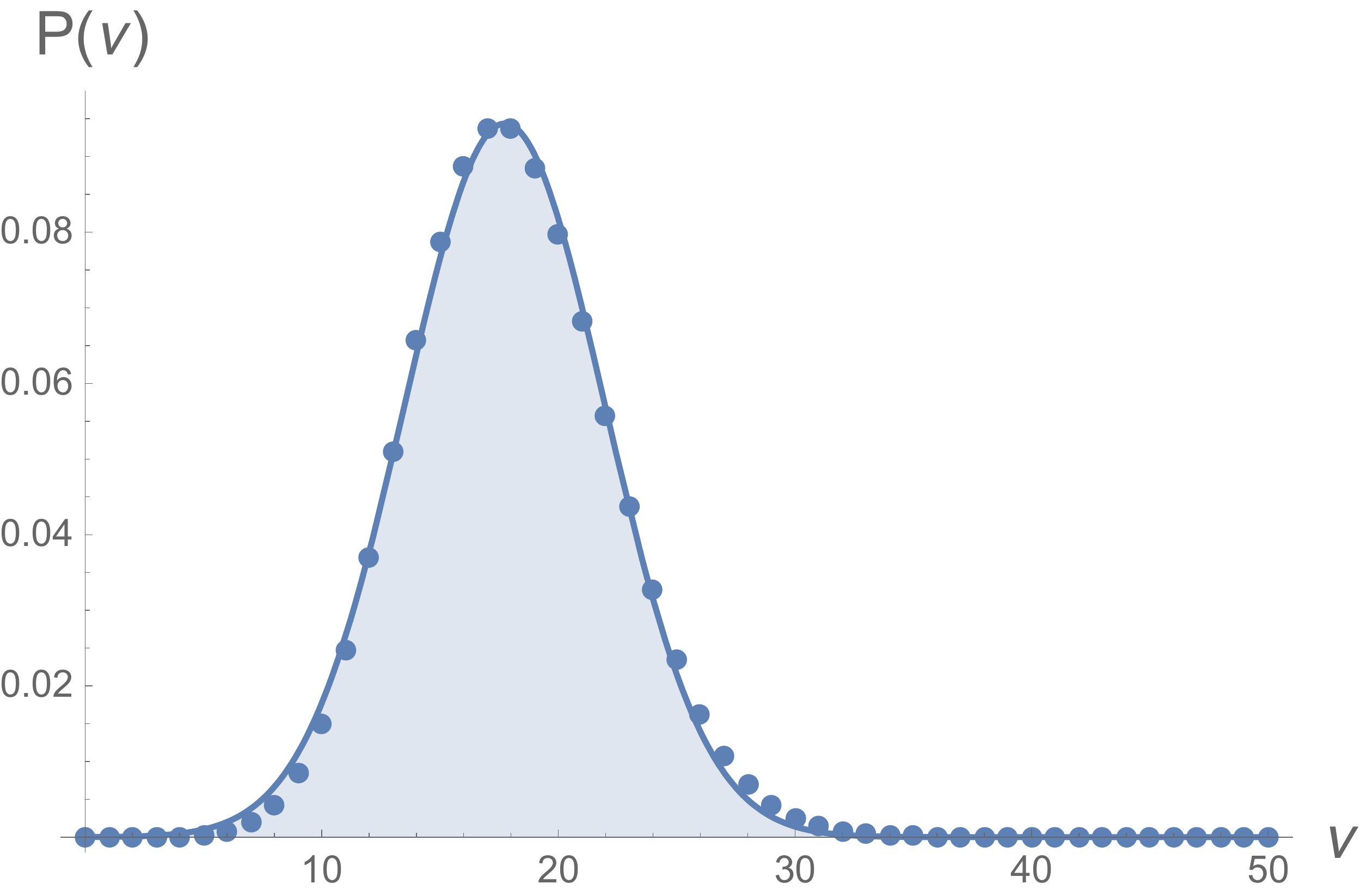}
\caption{%
For large values of the dipole interaction parameter $\mu$, all the
photon distributions (dots) are practically equal. They approach a
normal distribution (solid curve) centered around
$\overline{\nu}=17.74$ and standard deviation $\sigma=4.23$.  We
considered $N_{a}=2$ atoms and $\mu=3$.
\label{fig09}
}
\end{center}
\end{figure}

Fig.~\ref{fig07} compares the behaviour of the photon number
probability distribution for the coherent state with the even (left)
and odd (right) SACS states. As the coherent state contains both, the
even and odd number of excitations, this comparison is valid.  We can
see that the largest difference occurs as we approach $\mu=0.5$, the
phase transition from the normal regime to the collective one. For
large values of $\mu$ all the distributions are practically the same.
They fit a gaussian distribution centered around
$\overline{\nu}=17.74$ with standard deviation $\sigma=4.23$, as shown
in Fig.~\ref{fig09}, where $N_{a}=2$ atoms, and $\mu=3$.

In the coherent state approximation to the ground state we have always
a separable state. Using the SACS states approximation we have
entanglement between the atomic and field parts of the system.

To calculate the entanglement we write down the density matrix
corresponding to the state in Eq.~(\ref{eq038}) and the trace over the
electromagnetic part of the system to obtain the reduced density
operator of the atomic part. We obtain
   \begin{eqnarray}
      \fl
      &&\left(\mathbf{\rho}^{\pm}_{M}\right)_{n_{2}n_{3},n_{2}^{\prime}n_{3}^{\prime}}=
      \frac{N_{a}!\  \gamma_{2}^{n_{2}}\,\gamma_{2}^{\ast\ n_{2}^{\prime}}\,
      \gamma_{3}^{n_{3}}\,\gamma_{3}^{\ast\ n_{3}^{\prime}}} {\sqrt{(N_{a}-n_{2}-n_{3})!\,n_{2}!\,n_{3}!
      \,(N_{a}-n_{2}^{\prime}-n_{3}^{\prime})!\,n_{2}^{\prime}!\,n_{3}^{\prime}!}}\,
     \nonumber\\
      \fl
      &&\quad\times\frac{\Big[1+(-1)^{\lambda_{2}(n_{2}+n_{2}^{\prime})
      +\lambda_{3}(n_{3}+n_{3}^{\prime})}\Big]
      \Big[\exp(\left|\alpha\right|^{2})\pm(-1)^{\lambda_{2}n_{2}
      +\lambda_{3}n_{3}}\exp(-\left|\alpha\right|^{2})\Big]} {\Big[2\,\left(\exp(\left|\alpha\right|^{2})\,
      (\gamma^{\ast}\cdot\gamma)^{N_{a}}
      \pm\exp(-\left|\alpha\right|^{2})\,(\gamma^{\ast}\cdot\tilde{\gamma})^{N_{a}}
      \right)\Big]}\ .
   \end{eqnarray}

\begin{figure}[h]
\begin{center}
    \includegraphics[scale=0.25]{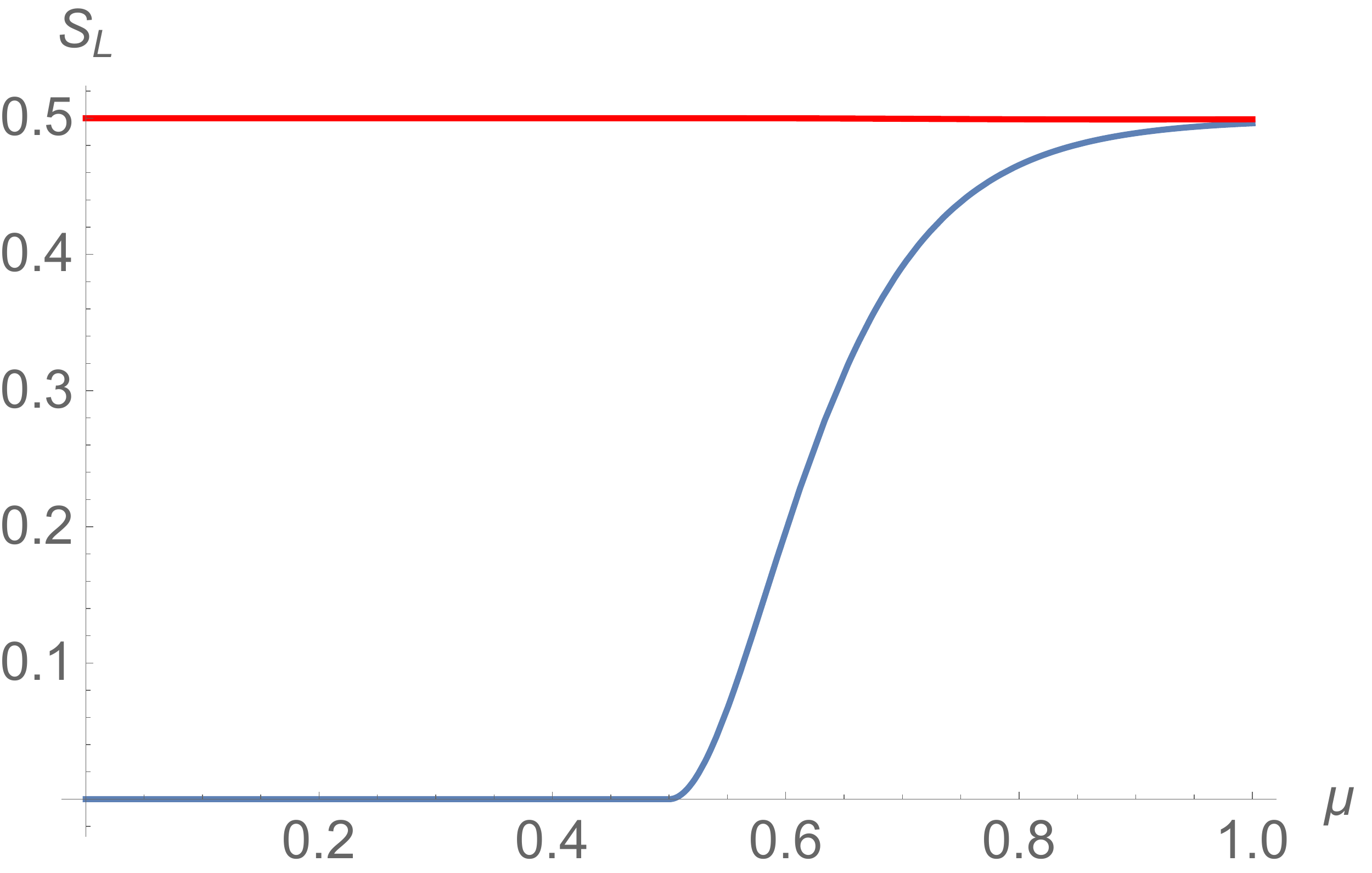}
\caption{%
(Colour online) Linear entropy for the even and odd SACS
approximations to the ground and first-excited states. We considered
$N_{a}=2$ atoms. The upper curve (red) corresponds to the odd SACS
state.
\label{fig10}
}
\end{center}
\end{figure}

The linear entropy, or purity, which gives a good measure of the
entanglement, is defined as
$S_{L}=1-\tr(\mathbf{\rho}^{\pm}_{M})^{2}$. Evaluating the previous
expression at the minima one gets
   \begin{equation}
      S^{^{\pm}}_L = \frac{\left( 1 - e^{\frac{N_a ( 16 \mu^4 -1)}{4
       \mu^2}} \right) \left( 1 - (2 \mu)^{4 N_a} \right)}{2
       \left(1 \pm (2 \mu)^{2 N_a} e^{\frac{N_a ( 8 \mu^4 -1)}{4
       \mu^2}}\right)^2}\ .
   \end{equation}
In Fig.~\ref{fig10} we show this quantity for the SACS even and odd
approximations. We must emphasize that for the coherent state
approximation the result is zero. In contrast, $S^{-}_{L}$ has
constant value of $1/2$ for any value of coupling parameter $\mu$. For
large values of $\mu$ or the number of particles one has
$S^\pm_L=1/2$.

\section{Conclusions}

The symmetries of the full Hamiltonian (not using the RWA
approximation) which describes the dipolar interaction of three-level
atoms with a one-mode radiation field lead to a parity conservation in
the total excitation number. By using as trial states a combination of
coherent states which preserve this symmetry, we obtain a better
approximation for the ground state than with the standard coherent
states, and a good estimation for the first excited level.  This is
made explicit by considering the $V$-configuration in the double
resonant case, where we calculate the minimum of the energy and the
expectation values of many important observables of the system such as
the number of photons, the atomic populations, the number of
excitations, and their corresponding fluctuations. For the first time,
we exhibit analytic expressions for all of these quantities. Also
for the first time we obtain an analytic description for the phase
diagram in parameter space, which distinguishes the normal and
collective regions, and which gives us all the quantum phase
transitions of the ground state from one region to the other as we
vary the interaction parameters (the matter-field coupling constants)
of the model, in functional form.

We also introduce the analogous of the $Q$-Mandel factor for the total
number of excitations, which better displays the difference between
the modified coherent states and the standard ones. The photon number
distribution is also obtained. With the SACS states we are able to
write down the reduced density matrix, allowing us to use the tools of
quantum information theory to calculate interesting properties of the
two lowest states of the system. As an example, we calculate the
linear entropy of system to estimate the entanglement between matter
and radiation.

\appendix

\section{Unitary transformation of $\mathbf{H}_{R}$}
\label{appA}

The tranformation of $\mathbf{H}_{R}$ is given by
   \begin{equation}
      \mathbf{H}_{R}(\theta)
      =\exp(-i\,\theta\,\mathbf{M})\,\mathbf{H}_{R}\,\exp(i\,\theta\,\mathbf{M})\ .
   \end{equation}
One can easily find that
   \begin{eqnarray*}
      \frac{d\phantom{\theta}}{d\theta}\,\mathbf{H}_{R}(\theta)&=&\mathbf{U}(\theta)\,
      \Big[\,\mathbf{M},\,\mathbf{H}_{R}\Big]\,\mathbf{U}^{\dagger}(\theta)\ ,\\
      \frac{d^{2}\phantom{\theta}}{d\theta^{2}}\,\mathbf{H}_{R}(\theta)&=&
      \mathbf{U}(\theta)\,
      \Big[\mathbf{M},\,\Big[\,\mathbf{M},\,\mathbf{H}_{R}\Big]\,\Big]
      \,\mathbf{U}^{\dagger}(\theta)
      =-4\,\mathbf{H}_{R}(\theta)\ .
   \end{eqnarray*}
From these results we deduce that
   \begin{equation}
      \mathbf{H}_{R}(\theta)=\cos(2\theta)\,\mathbf{H}_{R}
      +\frac{i}{2}\,\sin(2\theta)\,\Big[\,\mathbf{M},\,\mathbf{H}_{R}\,\Big]\ .
   \end{equation}
Therefore the transformation under $\mathbf{U}(\theta)$ leaves
$\mathbf{H}_{R}$ invariant if $\theta=\pi+2\pi\,n$,
$n\in\mathbb{Z}$, and the states that are invariant under this
transformation are given by
   \begin{equation}
      \vert\alpha;\,\gamma\}_{\pm}:=\left(\mathbf{1}
      \pm\exp(i\,\pi\,\mathbf{M})\right)\,\vert\alpha;\,\gamma\}
      =\vert\alpha;\,\gamma\}\pm\vert\-\alpha;\,\tilde{\gamma}\}\ ,
   \end{equation}
where
$\tilde{\gamma}=(1,\,(-1)^{\lambda_{2}}\gamma_{2},\,(-1)^{\lambda_{3}}\gamma_{3})$.

\section{Matrix elements in the SACS basis}
\label{appB}

In order to calculate the energy surface of the Hamiltonian, with and without the RWA approximation, we need the following matrix elements. For the one-body terms, we have
\vspace{2mm}
 \begin{eqnarray}
   	  \fl
      {}_{\pm}\{\alpha;\,\gamma\vert\,\mathbf{A}_{ii}\,\vert\alpha;\,\gamma\}_{\pm}
      &=&2\,N_{a}\,\left|\gamma_{i}\right|^{2}\,\Big[\exp(\left|\alpha\right|^{2})\,
      (\gamma^{\ast}\cdot\gamma)^{N_{a}-1}\nonumber\\
      \fl
      &&\ \pm(-1)^{\lambda_{i}}\,\exp(-\left|\alpha\right|^{2})
      \,(\gamma^{\ast}\cdot\tilde{\gamma})^{N_{a}-1}\Big]\, ,\\[3mm]
       \fl
      {}_{\pm}\{\alpha\,\gamma\vert\,\mathbf{a}^{\dagger}\mathbf{a}\,\vert\alpha\,\gamma\}_{\pm}&=&2\,
      \left|\alpha\right|^{2}\,\Big[\,\exp(|\alpha|^{2})\,(\gamma^{\ast}\cdot\gamma)^{N_{a}}
      \mp\exp(-|\alpha|^{2})\,(\gamma^{\ast}\cdot\tilde{\gamma})^{N_{a}}\Big]\ .
   \end{eqnarray}
   
For the interaction terms, we have
\vspace{2mm}
    \begin{eqnarray}
   	\fl
      &&{}_{\pm}\{\alpha;\,\gamma\vert\,\mathbf{A}_{ij}\,\mathbf{a}\,\vert\alpha;\,\gamma\}_{\pm}
      =N_{a}\,\alpha\,\gamma_{i}^{\ast}\,\gamma_{j}\,\left(1-(-1)^{\lambda_{i}+\lambda_{j}}
      \right) \nonumber\\
      \fl
      &&\qquad\qquad\quad\times\Big[\exp(\left|\alpha\right|^{2})\,(\gamma^{\ast}\cdot\gamma)^{N_{a}-1}
      \pm(-1)^{\lambda_{i}}\,\exp(-\left|\alpha\right|^{2})
      \,(\gamma^{\ast}\cdot\tilde{\gamma})^{N_{a}-1}\Big]\ ,
    \end{eqnarray}
      
    \begin{eqnarray}
 	\fl
      &&{}_{\pm}\{\alpha;\,\gamma\vert\,(\mathbf{A}_{ij}+\mathbf{A}_{ji})
      \,(\mathbf{a}+\mathbf{a}^{\dagger})\,\vert\alpha;\,\gamma\}_{\pm}
      =N_{a}\,(\alpha+\alpha^{\ast})\,(1-(-1)^{\lambda_{i}+\lambda_{j}}) \nonumber\\
      	\fl
      &&\qquad\qquad\qquad\times\Big[\,\exp(|\alpha|^{2})\,
      (\gamma_{i}^{\ast}\,\gamma_{j}+\gamma_{j}^{\ast}\,\gamma_{i})
      \,(\gamma^{\ast}\cdot\gamma)^{N_{a}-1} \nonumber\\
      	\fl
      &&\qquad\qquad\qquad\pm\exp(-|\alpha|^{2})\,((-1)^{\lambda_{i}}\,\gamma_{i}^{\ast}\,\gamma_{j}
      +(-1)^{\lambda_{j}}\,\gamma_{j}^{\ast}\,\gamma_{i})
      \,(\gamma^{\ast}\cdot\tilde{\gamma})^{N_{a}-1}\Big]\ .
   \end{eqnarray}
   
For the fluctuations in the atomic populations and in the number of photons we need
\vspace{2mm}
   \begin{eqnarray}
   	\fl
      &&{}_{\pm}\{\alpha;\,\gamma\vert\,\mathbf{A}_{ii}^{2}\,\vert\alpha;\,\gamma\}_{\pm}
      =\nonumber\\
      	\fl
      &&\qquad\qquad 2\,N_{a}\,\left|\gamma_{i}\right|^{2}\,\Big[\exp(\left|\alpha\right|^{2})\,\left(
      (\gamma^{\ast}\cdot\gamma)^{N_{a}-1}+(N_{a}-1)\,(\gamma^{\ast}\cdot\gamma)^{N_{a}-2}
      \,\left|\gamma_{i}\right|^{2}\right)\nonumber\\
      \fl
      &&\qquad\qquad\pm\exp(-\left|\alpha\right|^{2})\,\left((-1)^{\lambda_{i}}\,
      (\gamma^{\ast}\cdot\tilde{\gamma})^{N_{a}-1}+(N_{a}-1)\,
      (\gamma^{\ast}\cdot\tilde{\gamma})^{N_{a}-2}\,\left|\gamma_{i}\right|^{2}\right)\Big]\ ,
    \end{eqnarray}   
    
    \begin{eqnarray}
      \fl
      {}_{\pm}\{\alpha\,\gamma\vert\,\left(\mathbf{a}^{\dagger}\mathbf{a}\right)^{2}\,\vert\alpha\,\gamma\}_{\pm}
      &=&2\,
      |\alpha|^{2}\,\Big[\exp(|\alpha|^{2})\,\left(\left|\alpha\right|^{2}+1\right)
      \,(\gamma^{\ast}\cdot\gamma)^{N_{a}}\nonumber\\
      \fl
      &&\qquad\pm\exp(-|\alpha|^{2})\,\left(\left|\alpha\right|^{2}-1\right)\,(\gamma^{\ast}
      \cdot\tilde{\gamma})^{N_{a}}\Big]\ .
   \end{eqnarray}

 In order to see the atomic transitions from level $j$ to level $i$, and its quadratic form,
 \vspace{2mm}
    \begin{eqnarray}
   	\fl
      {}_{\pm}\{\alpha\,\gamma\vert\,\mathbf{A}_{ij}\,\vert\alpha\,\gamma\}_{\pm}&=&N_{a}\,\Big[
      \exp(\left|\alpha\right|^{2})\,\left(\gamma^{\ast}\cdot\gamma\right)^{N_{a}-1}\,
      \left(\gamma_{i}^{\ast}\,\gamma_{j}+\tilde{\gamma}_{i}^{\ast}\,\tilde{\gamma}_{j}
      \right)\nonumber\\
      \fl
      &&\quad\pm\exp(-\left|\alpha\right|^{2})\,\left(\gamma^{\ast}\cdot\tilde{\gamma}\right)^{N_{a}-1}\,
      \left(\gamma_{i}^{\ast}\,\tilde{\gamma}_{j}+\tilde{\gamma}_{i}^{\ast}\,\gamma_{j}\right)\Big]\ ,\\
      \fl
      {}_{\pm}\{\alpha\,\gamma\vert\,\mathbf{A}_{ij}\,\mathbf{A}_{kl}\,\vert\alpha\,\gamma\}_{\pm}&=&\Big[
      \exp(\left|\alpha\right|^{2})\,\Big(N_{a}\,\delta_{jk}\,(\gamma^{\ast}\cdot\gamma)^{N_{a}-1}\,
      \left(\gamma_{i}^{\ast}\,\gamma_{l}+\tilde{\gamma}_{i}^{\ast}\,\tilde{\gamma}_{l}\right)\nonumber\\
      \fl
      &+&N_{a}\,(N_{a}-1)\,(\gamma^{\ast}\cdot\gamma)^{N_{a}-2}\,\left(
      \gamma_{i}^{\ast}\,\gamma_{j}\,\gamma_{k}^{\ast}\,\gamma_{l}+
      \tilde{\gamma}_{i}^{\ast}\,\tilde{\gamma}_{j}\,\tilde{\gamma}_{k}^{\ast}\,\tilde{\gamma}_{l}\right)
      \Big)\nonumber\\
      \fl
      &\pm&\exp(-\left|\alpha\right|^{2})\,\Big(N_{a}\,\delta_{j\,k}\,(\gamma^{\ast}\cdot\tilde{\gamma})^{N_{a}-1}\,
      \left(\gamma_{i}^{\ast}\,\tilde{\gamma}_{l}+\tilde{\gamma}_{i}^{\ast}\,\gamma_{l}\right)\nonumber\\
      \fl
      &+&N_{a}\,(N_{a}-1)\,(\gamma^{\ast}\cdot\tilde{\gamma})^{N_{a}-2}\,\left(
      \gamma_{i}^{\ast}\,\tilde{\gamma}_{j}\,\gamma_{k}^{\ast}\,\tilde{\gamma}_{l}+
      \tilde{\gamma}_{i}^{\ast}\,\gamma_{j}\,\tilde{\gamma}_{k}^{\ast}\,\gamma_{l}\right)
      \Big)\Big]\ ;  
   \end{eqnarray}
these two expressions allow us to check the first and second order Casimir invariants of $U(3)$.

It is also interesting to evaluate the expectation value of the total number of excitations
$\mathbf{M}$, and that of its square in order to obtain the fluctuation and the equivalent to the Mandel parameter:
\vspace{2mm} 
   \begin{eqnarray}
   	 \fl
	 &&{}_{\pm}\{\alpha;\,\gamma\vert\,\mathbf{M}\,\vert\alpha;\,\gamma\}_{\pm}
	\nonumber\\
	\fl
	&&\qquad=2\,\exp(\left|\alpha\right|^{2})\,(\gamma^{\ast}\cdot\gamma)^{N_{a}-1}
	\,\Big[\left|\alpha\right|^{2}\,(\gamma^{\ast}\cdot\gamma)+N_{a}
	\sum_{i=2}^{3}\lambda_{i}\,\left|\gamma_{i}\right|^{2}\Big]\nonumber\\
	\fl
	&&\qquad\pm 2\,\exp(-\left|\alpha\right|^{2})\,(\gamma^{\ast}\cdot\tilde{\gamma})^{N_{a}-1}
	\,\Big[-\left|\alpha\right|^{2}\,(\gamma^{\ast}\cdot\tilde{\gamma})+N_{a}
	\sum_{i=2}^{3}(-1)^{\lambda_{i}}\,\lambda_{i}\,\left|\gamma_{i}\right|^{2}\Big]\ ,\\
   	 \fl
      &&{}_{\pm}\{\alpha;\,\gamma\vert\,\mathbf{M}^{2}\,\vert\alpha;\,\gamma\}_{\pm}
      =2\,\left|\alpha\right|^{2}\,\Big[\exp(\left|\alpha\right|^{2})\,\left(\left|\alpha\right|^{2}+1\right)
      (\gamma^{\ast}\cdot\gamma)^{N_{a}}\nonumber\\
      \fl
      &&\quad\qquad\pm\exp(-\left|\alpha\right|^{2})\,\left(\left|\alpha\right|^{2}-1\right)
      (\gamma^{\ast}\cdot\tilde{\gamma})^{N_{a}}\Big]\nonumber\\
      \fl
      &&\qquad+2\,\lambda_{2}^{2}\,N_{a}\,\left|\gamma_{2}\right|^{2}\Big[
      \exp(\left|\alpha\right|^{2})\,\left((N_{a}-1)\,\left|\gamma_{2}\right|^{2}\,
      (\gamma^{\ast}\cdot\gamma)^{N_{a}-2}+(\gamma^{\ast}\cdot\gamma)^{N_{a}-1}\right)
      \nonumber\\
      \fl
      &&\quad\qquad\pm\exp(-\left|\alpha\right|^{2})\,\left((N_{a}-1)\,\left|\gamma_{2}\right|^{2}\,
      (\gamma^{\ast}\cdot\tilde{\gamma})^{N_{a}-2}+(-1)^{\lambda_{2}}
      (\gamma^{\ast}\cdot\tilde{\gamma})^{N_{a}-1}\right)\Big]\nonumber\\
      \fl
      &&\qquad+2\,\lambda_{3}^{2}\,N_{a}\,\left|\gamma_{3}\right|^{2}\Big[
      \exp(\left|\alpha\right|^{2})\,\left((N_{a}-1)\,\left|\gamma_{3}\right|^{2}\,
      (\gamma^{\ast}\cdot\gamma)^{N_{a}-2}+(\gamma^{\ast}\cdot\gamma)^{N_{a}-1}\right)
      \nonumber\\
      \fl
      &&\quad\qquad\pm\exp(-\left|\alpha\right|^{2})\,\left((N_{a}-1)\,\left|\gamma_{3}\right|^{2}\,
      (\gamma^{\ast}\cdot\tilde{\gamma})^{N_{a}-2}+(-1)^{\lambda_{3}}
      (\gamma^{\ast}\cdot\tilde{\gamma})^{N_{a}-1}\right)\Big]\nonumber\\
      \fl
      &&\qquad+4\,N_{a}\,\lambda_{2}\,\left|\alpha\right|^{2}\,\left|\gamma_{2}\right|^{2}\Big[
      \exp(\left|\alpha\right|^{2})\,(\gamma^{\ast}\cdot\gamma)^{N_{a}-1}
      \mp(-1)^{\lambda_{2}}\exp(-\left|\alpha\right|^{2})\,
      (\gamma^{\ast}\cdot\tilde{\gamma})^{N_{a}-1}\Big]\nonumber\\
      \fl
      &&\qquad+4\,N_{a}\,\lambda_{3}\,\left|\alpha\right|^{2}\,\left|\gamma_{2}\right|^{2}\Big[
      \exp(\left|\alpha\right|^{2})\,(\gamma^{\ast}\cdot\gamma)^{N_{a}-1}
      \mp(-1)^{\lambda_{3}}\exp(-\left|\alpha\right|^{2})\,
      (\gamma^{\ast}\cdot\tilde{\gamma})^{N_{a}-1}\Big]\nonumber\\
      \fl
      &&\qquad+4\,N_{a}(N_{a}-1)\,\lambda_{2}\,\lambda_{3}\,\left|\gamma_{2}\right|^{2}\,
      \left|\gamma_{3}\right|^{2}\,\Big[
      \exp(\left|\alpha\right|^{2})\,(\gamma^{\ast}\cdot\gamma)^{N_{a}-2}\nonumber\\
      \fl
      &&\quad\qquad\mp(-1)^{\lambda_{2}+\lambda_{3}}\exp(-\left|\alpha\right|^{2})\,
      (\gamma^{\ast}\cdot\tilde{\gamma})^{N_{a}-2}\Big]\ .
   \end{eqnarray}

\vskip2.0pc

\section*{Acknowledgments}
This work was partially supported by CONACyT-M\'exico, and DGAPA-UNAM (under projects IN101614 and IN110114).

\end{document}